\documentclass[12pt]{iopart}

\usepackage[utf8]{inputenc}
\usepackage[english]{babel}

\usepackage{iopams}
\usepackage{setstack}

\usepackage{natbib}

\usepackage{booktabs}
\usepackage{hyperref}
\usepackage{orcidlink}
\usepackage{amssymb}

\usepackage{sidenotes}

\newcommand{\eqref}[1]{(\ref{#1})}

\newcommand{\N}{\mathbb{N}}
\newcommand{\R}{\mathbb{R}}

\newcommand{\defeq}{\overset{\mathrm{def}}{=\joinrel=}}

\newcommand{\Cc}{\mathcal{C}}

\newcommand{\Sc}{\mathcal{S}}
\newcommand{\Ss}{\mathsf{S}}
\newcommand{\Vc}{\mathcal{V}}
\newcommand{\dr}{\mathrm{d}}

\newcommand*{\transpose}{{\mkern-1.5mu\mathsf{T}}}

\newcommand{\eg}{\textit{e.g.}}
\newcommand{\ie}{\textit{i.e.}}

\makeatletter
\DeclareRobustCommand{\text}{%
  \ifmmode\expandafter\text@\else\expandafter\mbox\fi}
\let\nfss@text\text
\def\text@#1{{\mathchoice
  {\textdef@\displaystyle\f@size{#1}}%
  {\textdef@\textstyle\f@size{#1}}%
  {\textdef@\textstyle\sf@size{#1}}%
  {\textdef@\textstyle \ssf@size{#1}}%
  \check@mathfonts
  }%
}
\def\textdef@#1#2#3{\hbox{{%
                    \everymath{#1}%
                    \let\f@size#2\selectfont
                    #3}}}
\makeatother

\usepackage{array}
\newcolumntype{L}[1]{>{\raggedright\let\newline\\\arraybackslash\hspace{0pt}}m{#1}}
\newcolumntype{C}[1]{>{\centering\let\newline\\\arraybackslash\hspace{0pt}}m{#1}}
\newcolumntype{R}[1]{>{\raggedleft\let\newline\\\arraybackslash\hspace{0pt}}m{#1}}

\begin{document}

\title[Efficient Radiation Treatment Planning based on Voxel Importance]{Efficient Radiation Treatment Planning \\ based on Voxel Importance}

\author{Sebastian Mair$^{1,\star}$\orcidlink{0000-0003-2949-8781}, Anqi Fu$^2$\orcidlink{0000-0002-2876-2942} and Jens Sjölund$^1$\orcidlink{0000-0002-9099-3522}}

\address{$^1$ Uppsala University, Sweden}
\address{$^2$ Memorial Sloan-Kettering Cancer Center, New York, USA}
\address{$^\star$ Author to whom any correspondence should be addressed.}

\ead{sebastian.mair@it.uu.se, fua@mskcc.org, jens.sjolund@it.uu.se}

\begin{abstract}
\textit{Objective.}
Radiation treatment planning involves optimization over a large number of voxels, many of which carry limited information about the clinical problem.
We propose an approach to reduce the large optimization problem by only using a representative subset of informative voxels.
This way, we drastically improve planning efficiency while maintaining the plan quality.

\textit{Approach.}
Within an initial probing step, we pre-solve an easier optimization problem involving a simplified objective from which we derive an importance score per voxel.
This importance score is then turned into a sampling distribution, which allows us to subsample a small set of informative voxels using importance sampling.
By solving a -- now reduced -- version of the original optimization problem using this subset, we effectively reduce the problem's size and computational demands while accounting for regions where satisfactory dose deliveries are challenging.

\textit{Main results.}
In contrast to other stochastic (sub-)sampling methods, our technique only requires a single probing and sampling step to define a reduced optimization problem.
This problem can be efficiently solved using established solvers without the need of modifying or adapting them.
Empirical experiments on open benchmark data highlight substantially reduced optimization times, up to 50~times faster than the original ones, for intensity-modulated radiation therapy (IMRT), all while upholding plan quality comparable to traditional methods.

\textit{Significance.}
Our novel approach has the potential to significantly accelerate radiation treatment planning by addressing its inherent computational challenges.
We reduce the treatment planning time by reducing the size of the optimization problem rather than modifying and improving the optimization method.
Our efforts are thus complementary to many previous developments.
\end{abstract}

\vspace{2pc}

\vspace{28pt plus 10pt minus 18pt}
\noindent{\small\rm Published at: {\it Phys. Med. Biol.}\par}

\maketitle

\section{Introduction}
Radiation treatment planning (RTP) is widely used for cancer treatment, where the goal is to achieve tumor control by delivering a sufficiently high dose to the target while sparing healthy tissue to the maximum possible extent.
This is often formulated as a constrained optimization problem where potentially conflicting objectives, such as tumor control and normal tissue sparing, are combined as a weighted sum in the objective function. 
The constraints are usually dictated by the physics and the hardware but can also include clinical considerations such as the maximum dose in an organ at risk.
Since the problem is often high-dimensional regarding variables and constraints, it can be computationally demanding to solve.
What makes matters worse, however, is that the weights of the different objective terms are challenging to determine a priori and, therefore, require a human (clinician) in the loop \citep{miettinen1999nonlinear,romeijn2004unifying}.

Consequently, many manual iterations are often needed to find a satisfactory trade-off.
This has motivated the wealth of research on efficient optimization methods tailored to RTP.
We, too, aim to reduce the planning time in RTP.
Still, rather than adapting and improving the optimization \emph{solver}, we focus on reducing the \emph{size} of the optimization problem instead.
Our efforts are thus complementary to much of the previous developments.

Standard formulations of the optimization problem consider all voxels in the regions of interest as if the voxel doses could be controlled independently by optimizing the beamlet intensities.
On the contrary, it follows from elementary radiation physics that the absorbed dose is strongly correlated in space due to effects such as Compton scattering and pair production \citep{fippel1999fast}.
A critical insight behind our approach is that a cleverly chosen subset of the voxels effectively pins down the dose absorbed in other areas of relevance. 
Such a subsampled -- and thus reduced -- optimization problem can be solved much faster than the optimization problem defined on all voxels.
This immediately raises two questions.
First, how can we identify such a representative subset? 
Second, can it be done efficiently enough to reduce the overall solution time?

In this paper, we describe an approach that answers these questions affirmatively.
Specifically, we derive an importance sampling strategy that automatically identifies a representative subset of voxels by \emph{probing}, \ie, approximately solving a simplified version of the original optimization problem at hand.
Our results demonstrate that we can reduce the solution time of IMRT problems to less than 7\%, 1.6\%, and 47\% for \emph{prostate}, \emph{liver}, and \emph{head and neck} cases, respectively, with essentially no difference in plan quality.
\looseness=-1

Contrary to the stochastic optimization method of \citet{martin2007accelerating}, we subsample \emph{just once} to define an optimization problem with substantially fewer constraints, which can then be solved with the solver of choice (according to the rich literature on RTP, see, \eg, \cite{wachter2006implementation,aleman2010interior,unkelbach2015optimization,SCS16,osqp}) without having the need of modifying or adapting them.
This is somewhat similar to \citet{thieke2002acceleration} and \citet{sjolund2019linear}, both of which used a geometric approach to define a representative subset.
\looseness=-1

In contrast, our approach is entirely data-driven and thereby accounts for case-specific differences, paving the way for personalized medicine.
We believe that probing, owing to its simplicity and efficiency, presents an appealing approach to shorten treatment planning times while maintaining their quality.

\section{Method}

\subsection{Notation and problem setting}\label{sec:problem}

Consider a patient volume consisting of $n\in\N$ voxels and let $n_\text{b}\in\N$ be the number of beamlets.
Given a beam configuration, we can simulate the dose each body voxel receives using an appropriate dose calculation technique \citep{ahnesjo1999dose}.
We focus on fluence map optimization, where the goal is to find the beamlet intensities $x\in\R^{n_\text{b}}_{\geq 0}$ yielding specific voxel doses $y\in\R^n_{\geq 0}$ that satisfy some clinical objectives.
The relationship between beamlet intensities and voxel doses is approximately linear, and described by a dose influence matrix $A\in\R^{n \times {n_\text{b}}}_{\geq 0}$ such that $Ax=y$.
The non-negativity of $A$ and $x$ are due to physical constraints and naturally yield a non-negative dose vector~$y$.
We associate each voxel with a target dose $t\in\R^n_{\geq 0}$ that is prescribed on a voxel basis.
The set of voxels $\Vc=[n]\overset{\text{def}}{=}\{1,2,\ldots,n\}$ is commonly divided into a group of distinct non-overlapping structures representing the planning target volume (PTV), organs at risk (OARs), and other generic body parts (BDY).
The resulting optimization problem is of the form
\begin{eqnarray}\label{eq:opt}
\min_{x,y} \ f(y) \quad \mathrm{s.t.} \quad Ax=y, \quad  y \in \Cc, \quad x \geq 0,
\end{eqnarray}
where $\Cc$ denote some (usually affine) constraints, \eg, mean and max dose constraints.
The dose penalty function $f$ is usually convex and separable in the number of voxels, \ie,
\looseness=-1
\begin{eqnarray}\label{eq:obj}
f(y) = \sum_{s\in\Ss}\sum_{v\in\Vc_s} \frac{1}{|\Vc_s|} f_s(y_v).
\end{eqnarray}
Here, $s \in \Ss$ denotes a structure, $\Ss=\{\text{PTV}, \text{OAR}, \text{BDY}\}$ is the set of all structures, $v\in[n]$ is the index of a voxel, and $\Vc_s$ is the set of all voxels in structure~$s$.
The per-voxel penalty is often of a form such as \citep{romeijn2003novel,fu2019convex,sjolund2019linear}
\begin{eqnarray}\label{eq:penalty}
f_s(y_v) = \frac{1}{\max(t_s,\tau_s)^p} \Big ( w_s^- \cdot \underbrace{([y_v-t_v]^-)^p}_{\text{underdosing}} + w_s^+ \cdot \underbrace{([y_v-t_v]^+)^p}_{\text{overdosing}} \Big ),
\end{eqnarray}
where $p\in\{1,2\}$ is an exponent, $t_v$ denotes the target dose per voxel, $t_s$ indexes the vector of target doses per structure, $\tau_s>0$ is a threshold that prevents a division by zero, $y_v$ is the dose per voxel, and $w_s^-, w_s^+ \geq 0$ are structure-dependent weights that penalize underdosing and overdosing, respectively.
Here, we use the element-wise operators $[z]^-=\max(-z,0)$ and $[z]^+=\max(z,0)$.
Note that the leading division in Equation~\eqref{eq:penalty} normalizes the penalty by the corresponding target dose $t_s$, in addition to the normalization by the number of voxels per structure $|\Vc_s|$ in Equation \eqref{eq:obj}.
Therefore, we can expect that the contribution of each structure to the objective function value is on the order of one.\footnote{As a physicist might say, the objective function is non-dimensionalized.}
Often, organs at risk have a target dose of zero.
Thus, setting a threshold, \eg, $\tau_\text{OAR}=1$ prevents a division by zero.
If $\Cc$ is an affine set, then choosing $p=1$ within the per-voxel penalty $f_s$ in Equation~\eqref{eq:penalty} yields a resulting optimization problem in Equation~\eqref{eq:opt} that is a linear program (LP) while $p=2$ yields a quadratic program (QP). 

Since the per-voxel penalty in Equation~\eqref{eq:penalty} is a convex function, so is the problem in Equation~\eqref{eq:opt} as a whole, as long as $\Cc$ is convex. Nevertheless, it is non-trivial to optimize because of the large scale and the (nominal) non-smoothness of the per-voxel penalty.
Moreover, the individual weights $w_s^-$ and $w_s^+$ have to be chosen per structure~$s$, which results in another optimization problem.

\subsection{Efficiency through subsampling}\label{sec:efficiency}

We propose to restrict the computation of the treatment plan and, thus, the costly optimization problem to only use a \emph{representative subset} of voxels.
Thus, instead of using all $n$ voxels, we only use $m \ll n$ \emph{informative} voxels.
The computational savings in time are then amplified when a selection of the structure-dependent penalty weights $w_s^-$ and $w_s^+$ is performed.

The idea is as follows.
We first optimize a simple probing function that approximates the original problem from Equation~\eqref{eq:opt} but is much easier to optimize (Subsection~\ref{sec:probing}).
Based on the probing function, we derive a \emph{per-voxel score} that is indicative of a voxel's informativeness (Subsection~\ref{sec:scoring}).
Specifically, we expect that non-target voxels close to the PTV and non-target voxels
directly intersected/traversed by the beam are important.
We then turn the scores into a probability distribution over voxels and sample a subset of voxels (Subsection~\ref{sec:sampling}).
After eliminating voxels anticipated to have low impact via subsampling, we can solve all remaining optimization problems (Equation~\eqref{eq:opt}) faster and more efficiently using this smaller data set.

\subsection{Probing function}\label{sec:probing}

Consider a simplified per-voxel penalty that does not have any additional parameters and neither distinguishes between structures nor overdosing and underdosing, \ie, $f_s(y_v) = (y_v - t_v)^2$, where $y_v = A_{v,\colon} x$.
Here, $A_{v,\colon} \in \R^{n_\text{b}}$ refers to the row in $A$ belonging to voxel~$v$.
Plugging this simplified per-voxel penalty into the optimization problem in Equation~\eqref{eq:opt} and neglecting the dose constraints $y\in\Cc$ results in the simple non-negative least squares problem: 
\begin{eqnarray}\label{eq:probing}
\min_{x\geq 0} \ f_\text{probing}(x), \quad \text{where} \quad f_\text{probing}(x) = \sum_{s\in\Ss}\sum_{v\in\Vc_s} \frac{1}{|\Vc_s|} (A_{v,\colon} x - t_v)^2.
\end{eqnarray}

We optimize this simplified objective over \emph{all} voxels using $x_0=0$ as an initial point.
This initialization is reasonable since it yields a dose of zero for all voxels, which is already ideal for all non-PTV voxels.
The optimization procedure now has to adapt the configuration $x$ to reach the target dose on PTV voxels while keeping the doses of OAR and BDY voxels low.
We utilize projected gradient descent (PGD)\footnote{Note that a closed-form solution is impossible due to the non-negativity constraints.}, where the projection onto the non-negative orthant can be made by simply clipping the negative coordinates in $x$ to zero.
After only a few iterations of the form
\looseness=-1
\begin{eqnarray*}
x_{k+1} = \left[x_k - \eta \frac{\dr}{\dr x} f_\text{probing}(x_k) \right]^+, 
\end{eqnarray*}
where $\eta>0$ is a step-size, we should be able to quantify which voxels are important, \ie, by their contribution to the objective.

\subsection{Scoring voxels based on their importance}\label{sec:scoring}

To derive importance scores, we adopt an approach from \cite{paul2021deep} to our scenario.
The idea is as follows.
We start by optimizing a cheap surrogate optimization problem, \ie, the probing function in Equation~\eqref{eq:probing}.
Let $g_k(v)=\frac{\dr}{\dr x} \frac{1}{|\Vc_s|} f_s(y_v)$, where $y_v = A_{v,\colon} x$, denote the gradient of the probing function of voxel $v$ at iteration $k$ of projected gradient descent.
\cite{paul2021deep} show that the influence of a voxel~$v$ given the current beam intensities $x_k$ during optimization depends on its gradient norm, \ie, $\left\|g_{k}(v)\right\|$.
Phrased differently, voxels with a small gradient norm have a small influence on the overall per-voxel penalty.
Likewise, we expect a high influence of voxels that have a large gradient norm.
From an optimization point of view, any voxel that receives a dose different from its target dose does not only contribute positively to the objective function value but also yields a gradient with a positive norm.
Hence, the gradient norm serves as a measure of importance.

In summary, we define the importance score as $\chi_k(v) = \left\|g_k(v)\right\|$, which depends on the number of steps $k$ of PGD.
In practice, we only derive this score after optimizing the probing function in Equation~\ref{eq:probing}.
For completeness, we provide the derivation of the score in \ref{app:score}.

\subsection{Subsampling voxels}\label{sec:sampling}

We now derive a sampling strategy based on importance sampling and start by turning the scores into a sampling distribution $q$ by normalizing the importance scores, \ie, $q(v)=\frac{\chi(v)}{\sum_{v'}\chi(v')}$.
Using the sampling distribution $q$, we can now sample a subset $\Sc\subset\Vc$ of size~$m \ll n$ and re-weight the per-voxel penalty in Equation~\eqref{eq:penalty} for every chosen voxel by multiplying by $\frac{1}{m \cdot q(v)}$. 
This re-weighting of voxels is needed so that the -- now reduced -- objective function yields an unbiased estimate of the objective function on all voxels.
\looseness=-1

\begin{figure}
\centering
\includegraphics[width=0.65\textwidth]{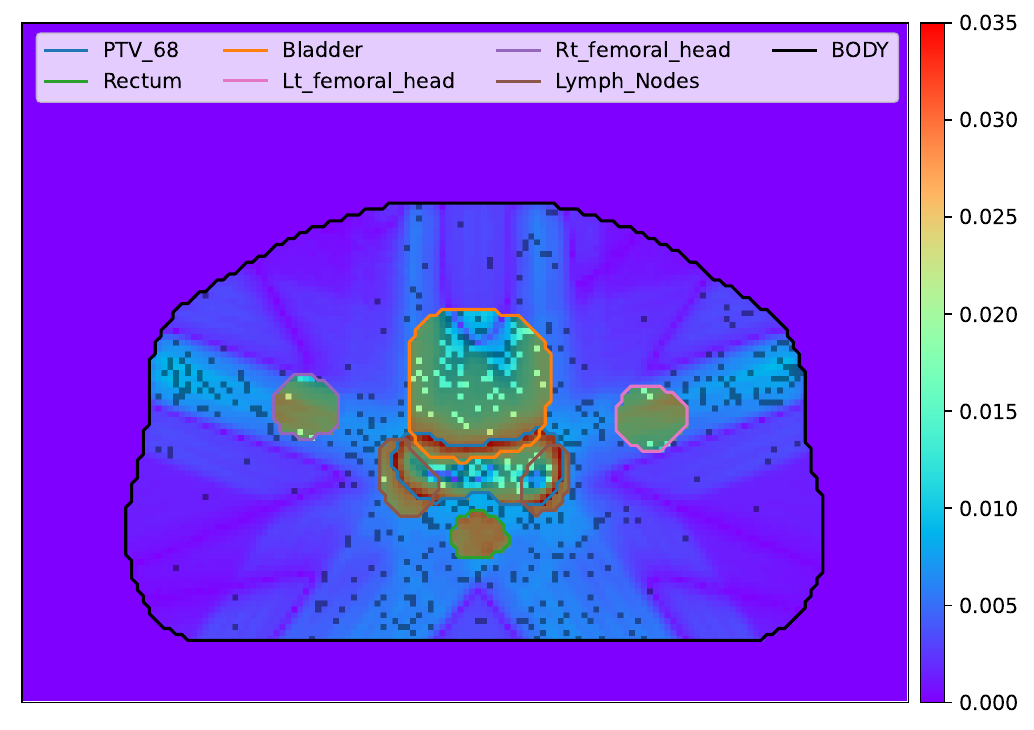}
\caption{Sampling distribution $q(v)$ for a prostate case including $m=51,778$ sampled voxels corresponding to 7.5\% of voxels.}
\label{fig:samples_prostate}
\end{figure}

An example of a sampling distribution and corresponding samples is shown in Figure~\ref{fig:samples_prostate} for a prostate case.
Note that BDY voxels are primarily sampled from the five radiation corridors.
Apart from BDY voxels, boundary areas between PTV and OAR are sampled more often.
The highest density is assigned to the boundary of the PTV.

\section{Results}
\subsection{Data}
In our experiments, we use the CORT data set \citep{craft2014shared}, which comes with four cases.
Neglecting the phantom case, we evaluate our proposed approach on a \emph{prostate}, \emph{liver}, and \emph{head and neck} case.
For all cases, we fix the couch angle to zero degrees, and we use five equally spaced gantry angles, \ie, 0, 72, 144, 216, and 288 degrees, for the prostate case.
For the liver case, we use six gantry angles, \ie, 32, 90, 148, 212, 270, and 328, and for the head and neck case, we use seven gantry angles, \ie, 0, 52, 104, 156, 208, 260, and 312.
More information on the data is provided in Table~\ref{tab:settings}.

\begin{table}[h]
\centering
\caption{Case specific information and settings}
\label{tab:settings}
\resizebox{.85\textwidth}{!}{
\begin{tabular}{lR{0.25\textwidth}R{0.2\textwidth}R{0.25\textwidth}}
\toprule
Information/Setting & Prostate & Liver & Head and Neck \\ 
\midrule
Number of PTV voxels & 6,770 & 6,954 & 5,768 \\ 
Number of OAR voxels & 26,469 & 148,318 & 2,162 \\ 
Number of BDY voxels & 657,134 & 1,772,085 & 243,963 \\ 
Number of beamlets $n_\text{b}$ & 721 & 389 & 7,906 \\ 
\midrule
$t_\text{PTV}$ & 68~Gy & 56~Gy & 70~Gy \\ 
$t_\text{OAR}$ and $t_\text{BDY}$ & 5~Gy & 5~Gy & 5~Gy \\ 
All $\tau_s$ & 5~Gy & 5~Gy & 5~Gy \\ 
\midrule
Chosen OAR structures & bladder, left and right femoral heads, lymph nodes, penile bulb, and rectum & heart, liver, spinal cord, and stomach & brain stem, chiasma, larynx, lips, left and right parotid, and spinal cord \\ 
Body structure & body & skin & external \\ 
\bottomrule
\end{tabular}
} 
\end{table}

\subsection{Setup}

All experiments run on an Intel Xeon machine with 28 cores with 2.60~GHz and 256~GB of memory and are implemented in Python using \texttt{numpy} \citep{harris2020array} and \texttt{CVXPY} \cite{diamond2016cvxpy}.
We compare our proposed subsampling approach (\texttt{gradnorm}) against uniform subsampling (\texttt{uniform}), \ie, using $\chi(v)=c$, where $c>0$ is an arbitrary constant, and the full solution (\texttt{full}), \ie, solving the optimization problem in Equation~\eqref{eq:opt} on all data.
For the solver, we use MOSEK\footnote{\url{https://www.mosek.com/products/version-10/}} with the same, default, settings for every run.
When optimizing the problem in Equation~\eqref{eq:opt}, we set the target doses for the OAR and BDY structures equal to $t_s = 5$ Gy.
This way, we penalize overdosing only for voxels that exceed a dose of~$t_s$.
For the probing function, we run projected gradient descent on Equation~\eqref{eq:probing} for 20 iterations from an initial configuration $x_0 = 0$ with a step size of $\eta = 2 (\sigma_{\min}(2A^\transpose A))+\sigma_{\max}(2A^\transpose A))^{-1}$.\footnote{Here, $\sigma_{\min}$ and $\sigma_{\max}$ denote the minimum and maximum singular values of the matrices they are applied to.}
Due to the stochasticity of the subsampling approaches, we repeat each experiment 50 times with different seeds and report on median performances.
In addition, we depict the 75\% and 25\% quartiles.
We use subsample sizes $m$ between 0.25\% and 90\% of the total number of voxels per case.
Unless stated otherwise, we present results for the choice of $p=2$, \ie, quadratic programming.
The weights $w_s^\pm$ were chosen such that \texttt{full} yields an acceptable dose-volume histogram (DVH).
Settings specific to the cases are described in Table~\ref{tab:settings}.

\begin{figure}
\includegraphics[width=0.49\textwidth]{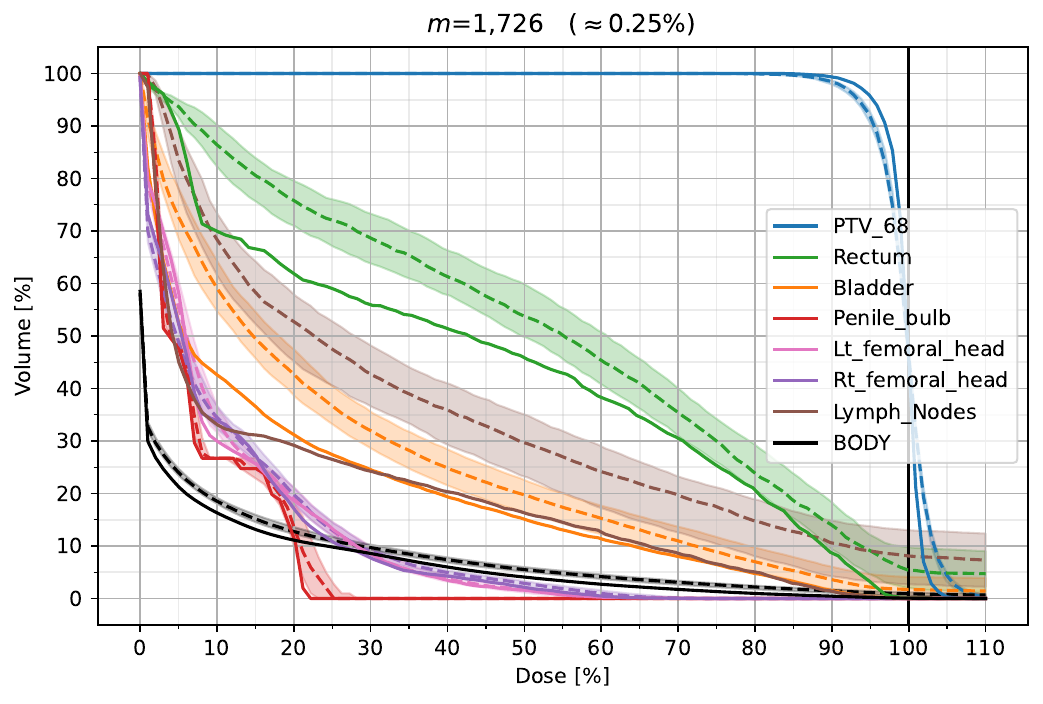}
\includegraphics[width=0.49\textwidth]{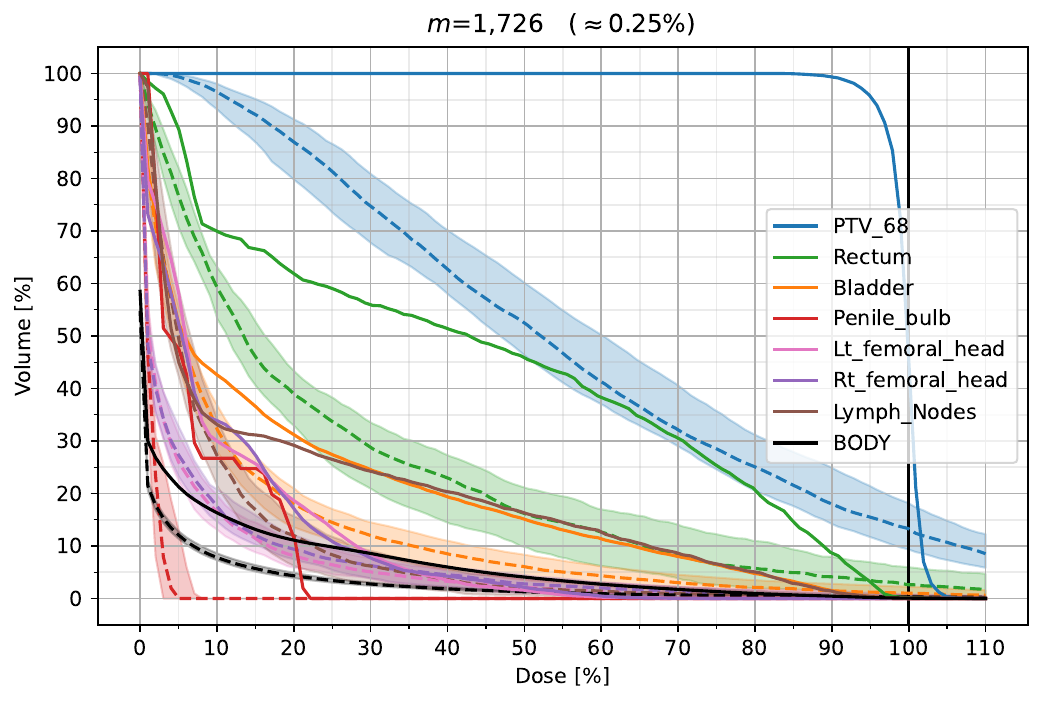}
\includegraphics[width=0.49\textwidth]{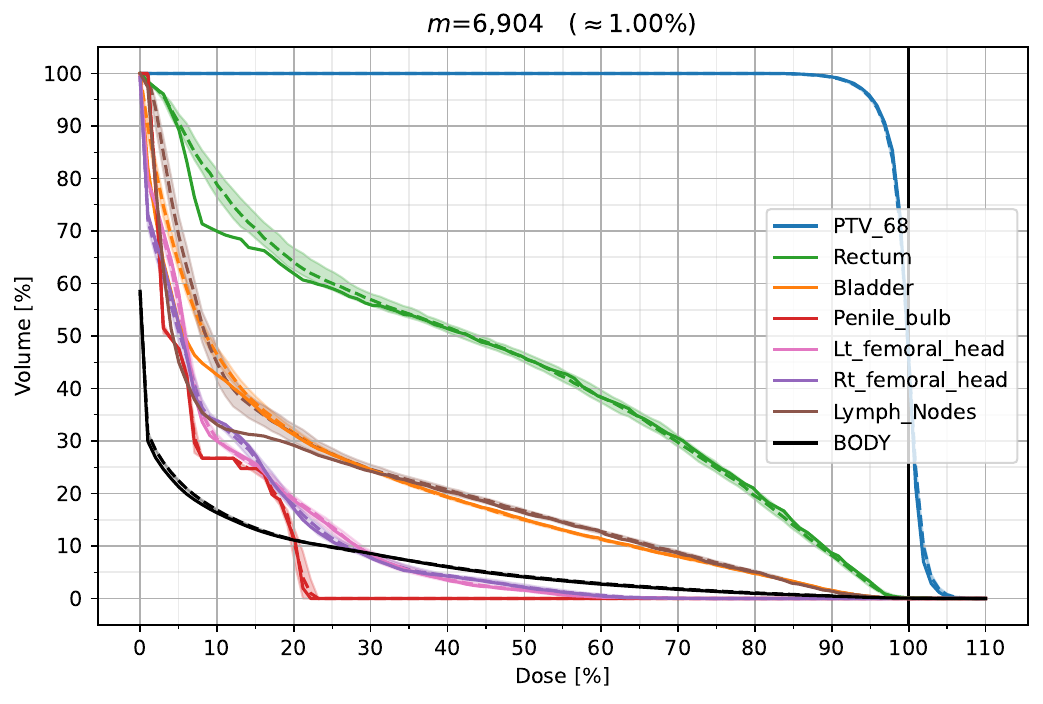}
\includegraphics[width=0.49\textwidth]{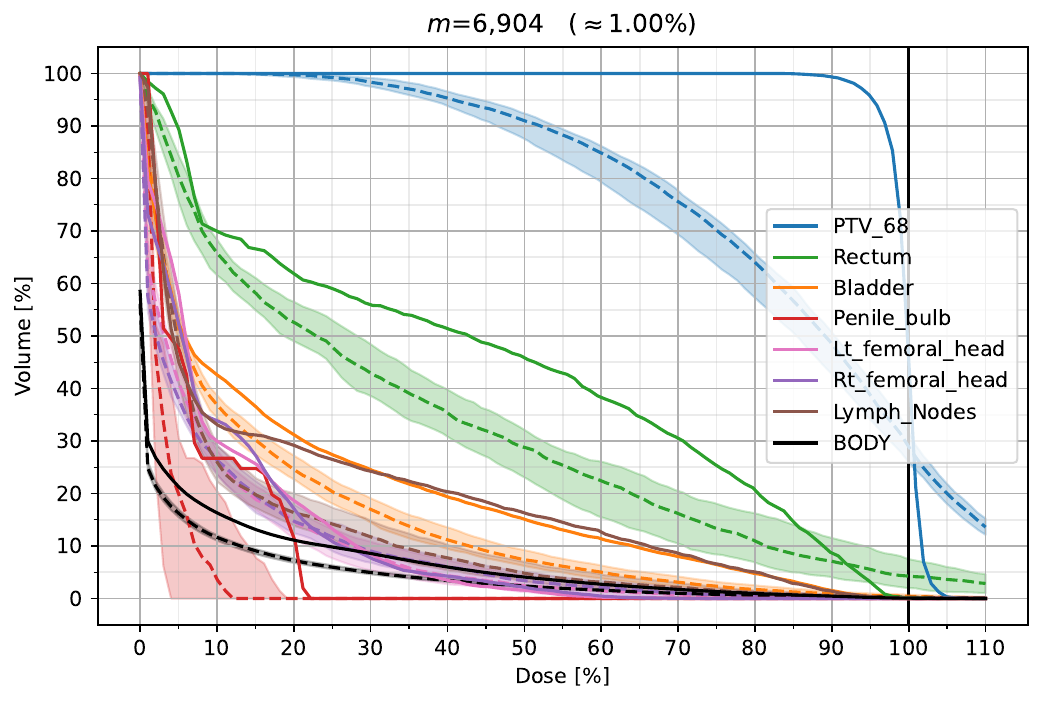}
\includegraphics[width=0.49\textwidth]{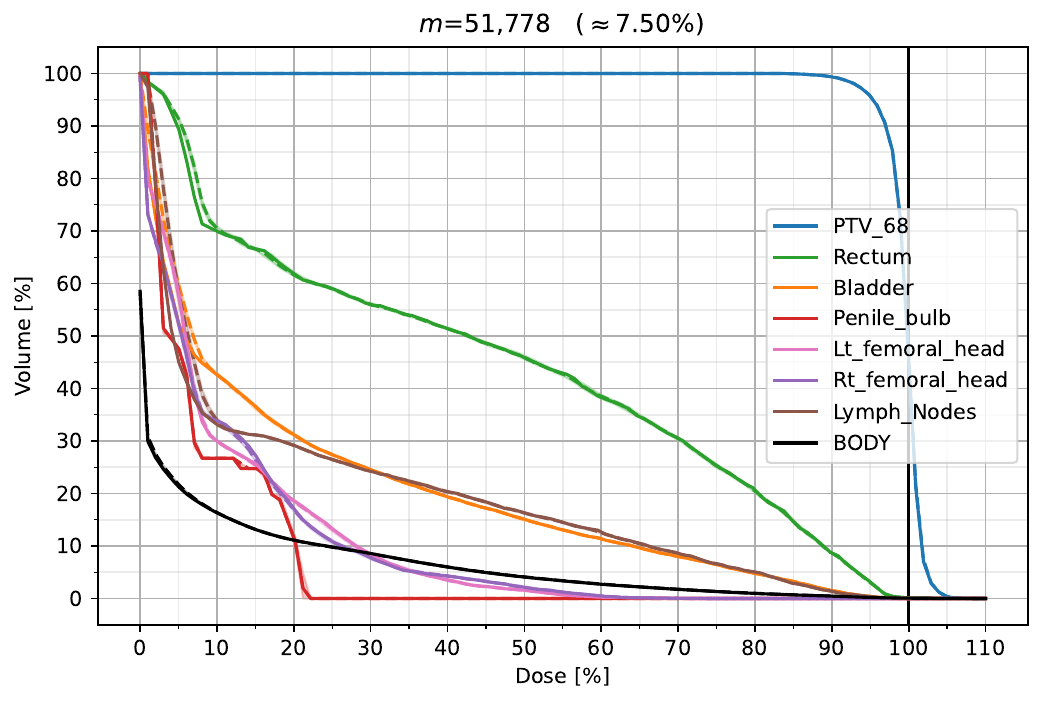}
\includegraphics[width=0.49\textwidth]{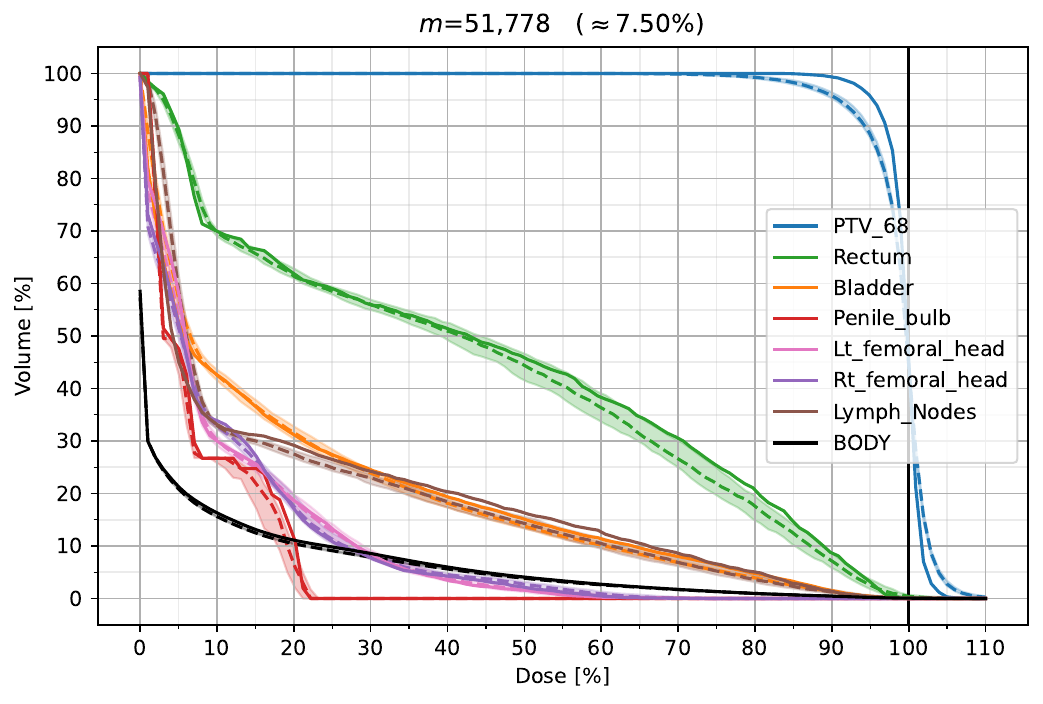}
\includegraphics[width=0.49\textwidth]{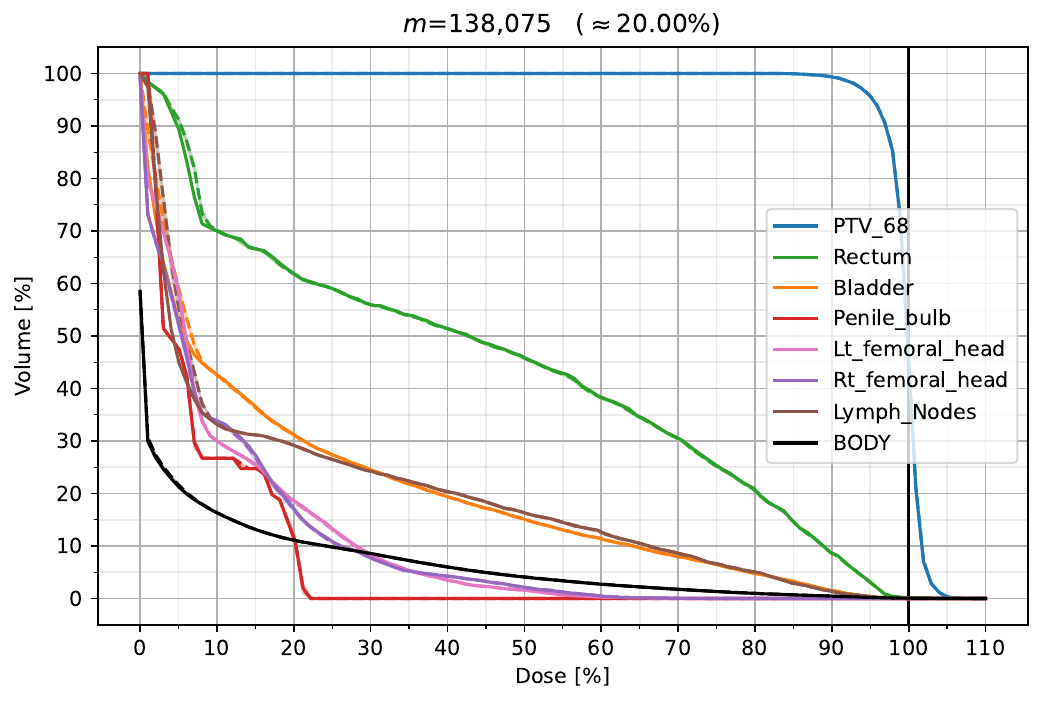}
\includegraphics[width=0.49\textwidth]{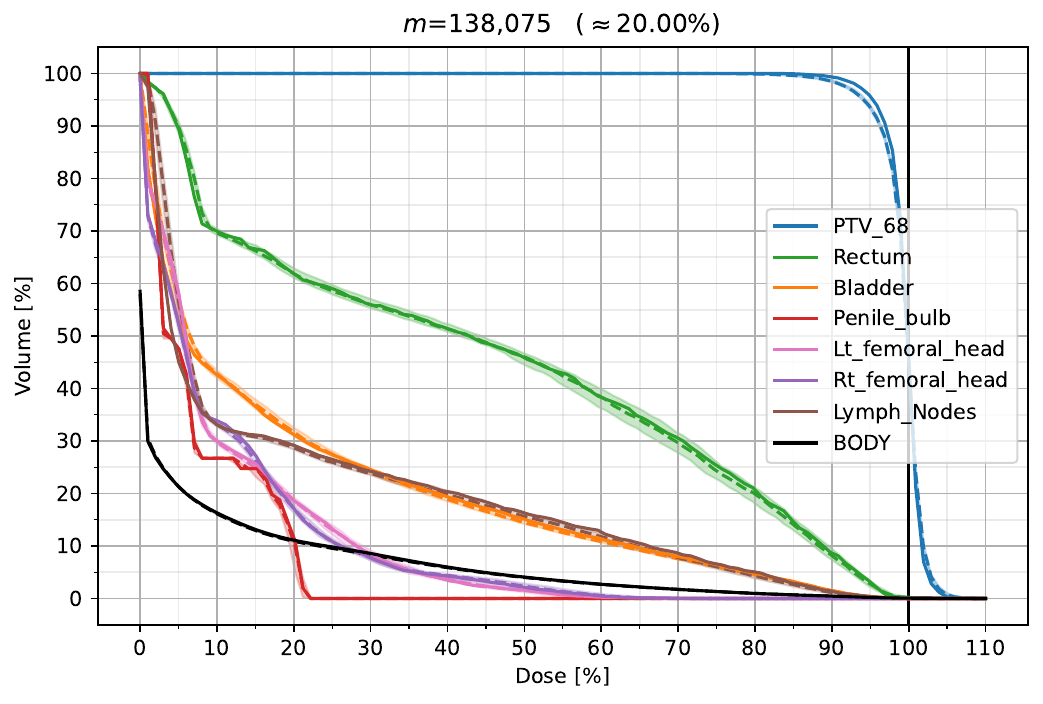}
\caption{Dose-volume histograms for the prostate case and $p=2$. Solid lines are the DVH from solving the problem with the full data set. The left column depicts the results of our proposed \texttt{gradnorm} subsampling approach while the right column shows \texttt{uniform} subsampling. The rows depict different subsampling sizes $m$ and the percentage of~$m$ relative to the total number of voxels $n$.}
\label{fig:DVHdist_prostate}
\end{figure}

\subsection{Evaluating subsampling approaches using dose-volume histograms}

\begin{figure}[t]
\centering
\includegraphics[width=0.49\textwidth]{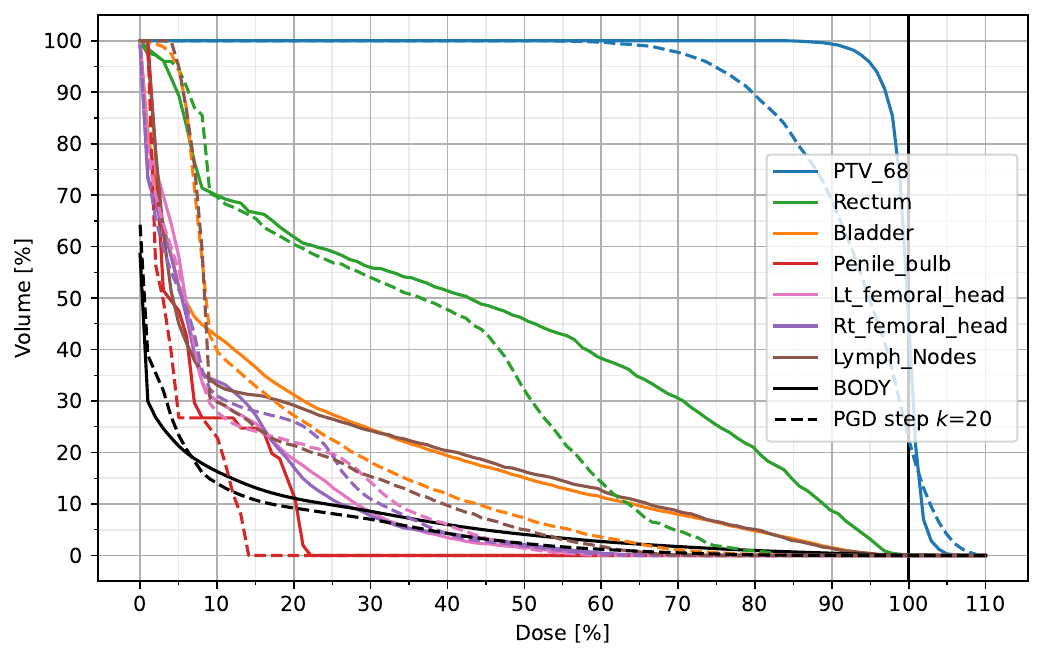}
\hfill
\includegraphics[width=0.49\textwidth]{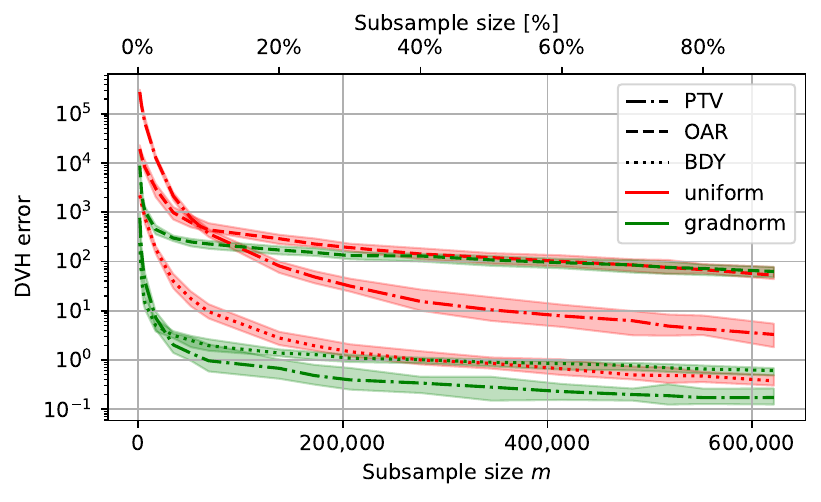}
\caption{Left: Dose-volume histograms for the prostate case of \texttt{full} (solid lines) and the projected gradient descent (PGD) optimization after $k=20$ steps (dashed lines) which we use as a probing function to derive the importance scores.
Right: Error of dose-volume histogram approximations of \texttt{uniform} and \texttt{gradnorm} when compared to \texttt{full} for the prostate case. The error is split among the structures $\Ss=\{\text{PTV}, \text{OAR}, \text{BDY}\}$.}
\label{fig:DVH}
\end{figure}

We now consider the prostate case, evaluate different subsampling sizes for \texttt{gradnorm}, and compare the performance in terms of DVHs against \texttt{uniform} and \texttt{full}.
Figure~\ref{fig:DVH} (left) depicts the DVH of \texttt{full} (solid lines) and the optimization of Equation~\eqref{eq:probing} using projected gradient descent (PGD) after $k=20$ steps (dashed lines).
The solution $x_{20}$ after 20 PGD steps yields a sub-optimal yet already informative solution and is used to derive the importance scores.
The resulting sampling distribution for a specific slice is depicted in Figure~\ref{fig:samples_prostate}.
\looseness=-1

After subsampling voxels according to the \texttt{gradnorm} and \texttt{uniform} strategies, and thus reducing the size of the optimization problem in Equation~\eqref{eq:opt}, we can evaluate their performances.
Figure~\ref{fig:DVHdist_prostate} depicts the results for \texttt{gradnorm} and \texttt{uniform} on the left and right columns, respectively.
Furthermore, the rows show various interesting subsampling sizes.
In all subplots, the medians of the subsampling approaches are depicted as dashed lines, while \texttt{full} is shown as solid lines.
The shaded areas reflect the 75\% and 25\% quartiles of all 50 repetitions.

We can see that \texttt{gradnorm} already approximates the PTV very well by using only 0.25\% of the voxels, while \texttt{uniform} needs more than 20\% to achieve a similar approximation.
Besides, almost all OARs are sufficiently well approximated using \texttt{gradnorm} with only 7.5\% of all voxels.
Note that the larger the subsample size $m$, the smaller the uncertainty.
The results for the liver and head and neck cases are deferred to \ref{app:results}.

In Figure~\ref{fig:DVH} (right), we aggregate the results shown in Figure~\ref{fig:DVHdist_prostate}.
Here, we compute the DVH error as a squared error between the DVH lines of the subsample approaches and \texttt{full} at 111 positions (from 0\% to 110\%).
As seen in the figure, regardless of the subsample size, \texttt{gradnorm} yields lower errors than \texttt{uniform} for all structures.

\subsection{Evaluating subsampling approaches using the objective function}

\begin{figure}[b]
\centering
\includegraphics[width=0.49\textwidth]{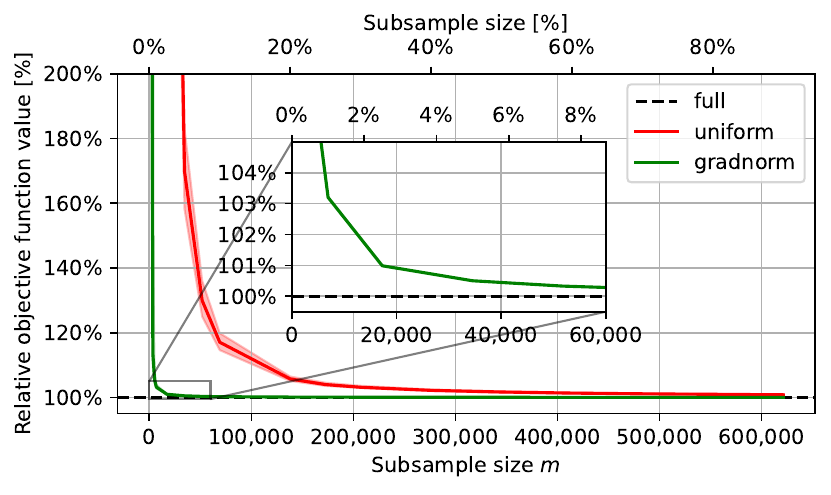}
\hfill
\includegraphics[width=0.49\textwidth]{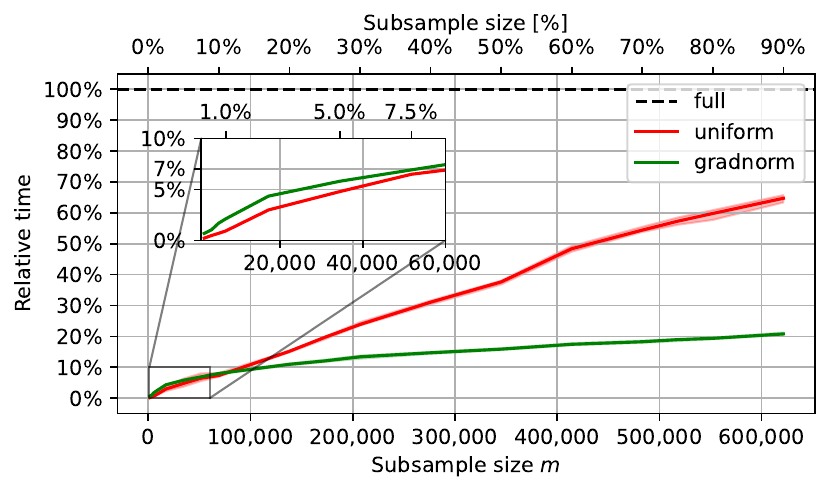}
\caption{Relative objective function values (left) and computation time (right) of \texttt{uniform} and \texttt{gradnorm} when compared to \texttt{full} for the prostate case.}
\label{fig:obj_reltime_subsample}
\end{figure}

Next, we evaluate the relative objective function value.
That is, the objective function value of the solution obtained from the reduced problem is divided by the objective function value of the solution obtained from the full problem.
The results are depicted in Figure~\ref{fig:obj_reltime_subsample} (left).
Once again, our \texttt{gradnorm} yields very quickly much better approximations than \texttt{uniform}.
Note that using 7.5\% of the data yields an objective function value that deviates less than 1\% of the corresponding one from the full problem.

\subsection{Evaluating subsampling approaches in terms of computational time}

Finally, we evaluate the time it takes to compute the radiation plans on the reduced optimization problems.
Figure~\ref{fig:obj_reltime_subsample} (right) depicts the relative computation times of \texttt{uniform} and \texttt{gradnorm} when compared to \texttt{full} for all three cases.
Note that \texttt{gradnorm} reduces the computation time to only approximately 20\% of the original time, regardless of the subsample size.
Specifically, using only 7.5\% of all voxels for \texttt{gradnorm} reduces the computation time to less than 7\%.

\section{Discussion}

When it comes to comparing the proposed subsampling strategy from Subsection~\ref{sec:sampling} to uniform subsampling, we can see from Figure~\ref{fig:samples_prostate} that our approach indeed leverages information, \eg, about the beam directions, obtained from the probing function.
Thus, our strategy can be seen as being \emph{data-driven}.
On the contrary, the uniform subsampling strategy does not have such additional information, and the user would need to adapt the strategy to include this information.
Note that in our proposed subsampling strategy, important voxels are sampled more often and are assigned a lower weight.
In contrast, less important voxels get sampled less often but are assigned a higher weight.

Since our strategy focuses on important or informative regions, it happens that we sample the same voxel multiple times.
Thus, the same penalty term appears multiple times in Equation~\eqref{eq:obj}, potentially leading to problems during optimization.
However, we can simply aggregate those terms by summing their weights.
A positive side-effect of this strategy is that the computation time does not necessarily increase at the same rate as the subsample size, see Figure~\ref{fig:obj_reltime_subsample} (right).

Note that the green line for \texttt{gradnorm} in Figure~4~(right) does not include the time for probing, \ie, the time to obtain the scores.
For example, to provide specific numbers, solving the optimization problem in Equation~\eqref{eq:opt} on all voxels of the prostate case takes approx. 289 seconds, while probing takes approx. 23.4 seconds (1.17 seconds per projected gradient descent iteration), and solving the -- now reduced -- optimization problem in Equation~\eqref{eq:opt} on only 7.5\% of informative voxels takes only approx. 16.8 seconds.
Note that the probing has to be done \emph{just once}, and any further optimization of Equation~\eqref{eq:opt} only costs an additional 16.8 seconds compared to the 289 seconds it takes to solve the problem on all voxels.
This is especially beneficial when the underdosing and overdosing penalty weights~$w_s^\pm$ per structure $s$ have to be optimized as well, resulting in a bilevel optimization task.  

Our approach is inspired by \cite{paul2021deep} but differs in several aspects.
First, \cite{paul2021deep} consider a neural network-based classifier while we use a (weighted) non-negative linear regression.
Second, they take the $m$ points with the largest scores while we turn the scores into a sampling distribution and subsample the set of voxels.
Third, their score is the expectation over several random initializations~$x_0$ while we only consider a single deterministic initialization, \ie, $x_0=0$.
We consider this adaptation reasonable since it initializes with a feasible point, and we optimize a convex function instead of a highly non-convex one, as is usual in deep learning.

Although we focus on IMRT optimization, we firmly believe that our proposed approach is also applicable to VMAT optimization and will lead to reduced optimization times, although not as drastically as in IMRT.
We see the evaluation for VMAT optimization as potential future research.

\subsection{Related work}

\cite{paul2021deep} were not the first to consider gradient norms as measures of informativeness in neural network training, as similar ideas were already leveraged by, \eg, \cite{alain2016variance} and \cite{katharopoulos2018not}.

Regarding radiation treatment planning, several performance-improving approaches have been proposed. 
\cite{thieke2002acceleration} sample pencil beams with a probability based on the radial distance between its central ray and a particular voxel.
They are able to reduce the dose-influence matrix to a third of its original size without impacting the quality of the resulting treatment plan. 
\cite{yang2004inverse} use an iterative algorithm to solve a weighted least squares problem, where each voxel's weight is updated based on the current dose of that voxel.
For a target voxel, the weight is negatively correlated with the dose, while for an OAR voxel, the weight is positively correlated.
The update equation depends on structure-specific parameters, which are found via trial and error.
\cite{zakarian2004wavelet} review several strategies for increasing the speed and efficiency of dose calculations, including a method based on fast wavelet transforms.
\cite{scherrer2005imrt} employ adaptive hierarchical clustering to reduce the size of the treatment planning problem.
Starting with a coarse clustering of voxels, they replace each cluster with the mean of its component voxels, solve the optimization problem using the mean data, and locally refine the clusters based on this solution.
The process continues until the solution is sufficiently close to that of the original problem, based on a derived error bound.
Numerical experiments showed that their adaptive clustering reduced the computation time substantially.
\cite{martin2007accelerating} formulate the treatment planning problem as an optimization problem with a sum of separable objectives.
To solve it, they employ stochastic steepest descent with voxel sampling in each iteration.
The authors compare the results from uniform sampling, manual sampling of each structure, and objective-specific sampling rates chosen by minimizing the variance of the total estimated objective.
They find that the objective-specific rates provide the largest speed improvement compared to regular steepest descent.
\cite{sjolund2019linear} perform sector-duration optimization for stereotactic radiosurgery by solving a linear programming problem.
As this problem is large, they reduce the solution time by uniformly sampling positions at random in each structure and solving the associated dual problem, resulting in time savings.
\cite{fu2019convex} tackle the treatment planning problem with dose-volume constraints.
This problem is a computationally challenging mixed-integer program.
To arrive at a solution, they first solve a convex approximation of the problem, then use the result to select a subset of voxels to dose bound so the dose-volume constraints are met with the smallest margin.
\cite{ungun2019cluster} solve a compressed approximation of the treatment planning problem by clustering voxels and beams with $k$-means clustering.
Using duality theory, they bound the optimality gap between the approximate and true solution.
With the proper choice of clusters, they are able to attain a substantial speedup while still producing an excellent treatment plan.

\section{Conclusion}
In this paper, we introduce the idea of reducing the size of the optimization problem within radiation treatment planning while maintaining the plan quality.
This is achieved by subsampling the large set of voxels of different information quality via importance sampling.
To do so, we derive a sampling distribution based on a score that reflects the \emph{informativeness} of a voxel.
This score is gathered by so-called \emph{probing}, \ie, by first pre-solving a simplified optimization problem and then using the gradient norm to measure a voxel's importance.
Using this subsampling technique allows us to significantly reduce the number of voxels, thus obtaining a much smaller optimization problem, which we can solve much faster.
A key advantage is that this reduced optimization problem can be solved using established solvers without modifying or adapting them.

Empirical evaluations on three openly available cases showcase the efficacy of our proposed approach.
For example, in a prostate case, using only 7.5\% of voxels yields deviations of the objective function on all voxels of less than 1\% while only taking less than 7\% of the time to compute.
In summary, our approach has the potential to significantly accelerate planning times while maintaining the plan quality.

\section*{Acknowledgments}

This work was partially supported by the Wallenberg AI, Autonomous Systems and Software Program (WASP) funded by the Knut and Alice Wallenberg Foundation; as well as Sweden's Innovation Agency (Vinnova) project 2022-03023.
The authors thank the anonymous reviewers for valuable comments on earlier drafts.

\section*{Data availability statement}

The data that support the findings of this study are openly available.
More specifically, we use the publicly available CORT data set from \cite{craft2014shared}.

\section*{Conflicts of interest}
The authors declare no conflict of interest.

\section*{Author contributions}
%

\textbf{Sebastian Mair:} Conceptualization, Formal Analysis, Investigation, Methodology, Software, Visualization, Writing - Original Draft.
\textbf{Anqi Fu:} Validation, Writing - Review and Editing.
\textbf{Jens Sjölund:} Conceptualization, Funding Acquisition, Project Administration, Validation, Writing - Review and Editing.

\section*{ORCID iDs}

Sebastian Mair \orcidlink{0000-0003-2949-8781} \url{https://orcid.org/0000-0003-2949-8781}

\noindent Anqi Fu \orcidlink{0000-0002-2876-2942} \url{https://orcid.org/0000-0002-2876-2942}

\noindent Jens Sjölund \orcidlink{0000-0002-9099-3522} \url{https://orcid.org/0000-0002-9099-3522}

\appendix

\section{Deriving the importance score}\label{app:score}
To derive importance scores, we adapt an approach from \cite{paul2021deep} to our scenario.
The idea is as follows.
We start by optimizing a cheap surrogate optimization problem, \ie, the probing function in Equation~\eqref{eq:probing}.
Let $\tilde{f}_s(y_v) \overset{\text{def}}{=} \frac{1}{|\Vc_s|} f_s(y_v)$ and $g_k(v)=\frac{\dr}{\dr x} \tilde{f}_s(y_v)$ denote the gradient of the probing function of voxel $v$ at iteration $k$ of projected gradient descent.
For the sake of the argument, assume that the training dynamics are in continuous time instead of discrete iterations.
The time derivative of the probing penalty for an arbitrary voxel $v$ given the set of all voxels $\Vc$ is
\begin{eqnarray*}
\Delta_k(v, \Vc)
&\defeq - \frac{\dr \tilde{f}_s(y_v)}{\dr k} 
 = - \frac{\dr \tilde{f}_s(y_v)}{\dr x_k} \frac{\dr x_k}{\dr k} 
 = - g_k(v) \frac{\dr x_k}{\dr k},
\end{eqnarray*}
where we first use the chain rule and then the definition of the function $g_k(v)$.
Note that because of
\begin{eqnarray*}
\frac{\dr x_k}{\dr k}
\approx x_{k+1} - x_{k}
= - \eta \sum_{v\in\Vc} g_k(v),
\end{eqnarray*}
we can simplify $\Delta$ to
\begin{eqnarray*}
\Delta_k(v, \Vc)
= \eta g_k(v) \sum_{v\in\Vc} g_k(v).
\end{eqnarray*}
Adapting an argument from \citet{paul2021deep}, we can bound the contribution of an arbitrary voxel $v$ as follows.
Let $\Vc_{\neg j}$ be the set of all voxels except the $j$th one.
Then, for any voxel $v$ the influence of having the $j$th voxel included during training can be bounded via
\begin{eqnarray*}
\|\Delta_k(v, \Vc) - \Delta_k(v, \Vc_{\neg j})\| 
&= \left\|\eta g(v) \sum_{v\in\Vc} g_k(v) - \eta g(v) \sum_{v\in\Vc_{\neg j}} g_k(v)\right\| \\ 
&= \left\|\eta g(v) g_k(v=j)\right\| \\ 
&\leq \left\|\eta g(v)\right\| \cdot \left\|g_k(v=j)\right\|.
\end{eqnarray*}
Since $\left\|\eta g(v)\right\|$ is a constant for all $j$, it can be neglected.
Thus, the influence of a voxel~$v$ given the current beam intensities $x_k$ during optimization depends on its gradient norm.
Phrased differently, voxels with a small gradient norm have a small influence on the overall per-voxel penalty.
Likewise, we expect a high influence of voxels that have a large gradient norm.
Hence, we define the importance score as $\chi_k(v) = \left\|g_k(v)\right\|$.

\section{Further results}\label{app:results}

\begin{figure}[h]
\centering
\includegraphics[height=6.75cm]{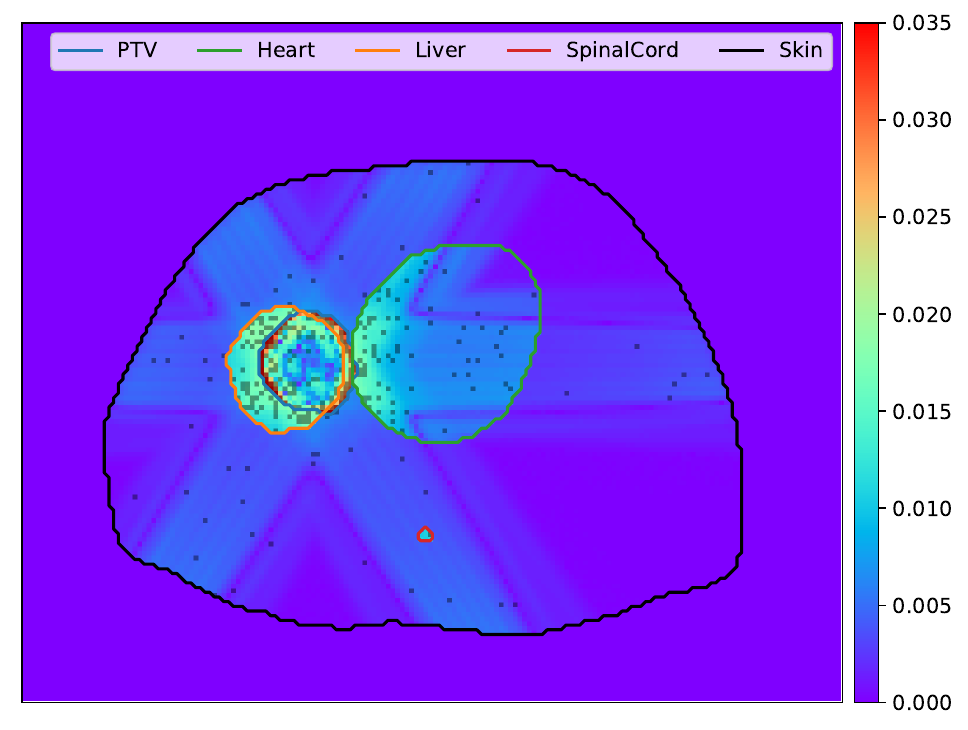}
\hfill
\includegraphics[height=6.75cm]{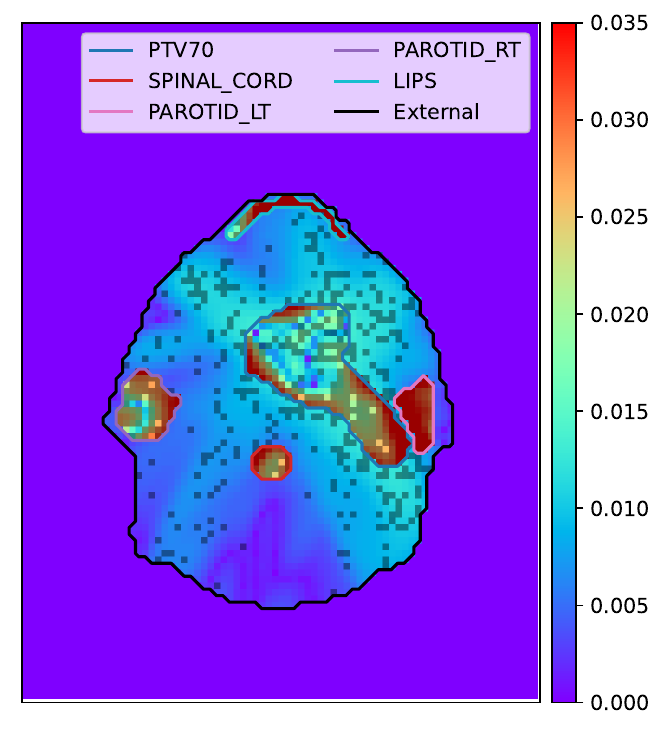}
\caption{Sampling distribution $q(v)$ for a liver (left) and head and neck case (right) including $m=9,637$ and $m=25,189$ sampled voxels corresponding to 0.5\% and 10\% of voxels, respectively.}
\label{app:fig:samples_rest}
\end{figure}

\begin{figure}[h]
\centering
\includegraphics[width=0.49\textwidth]{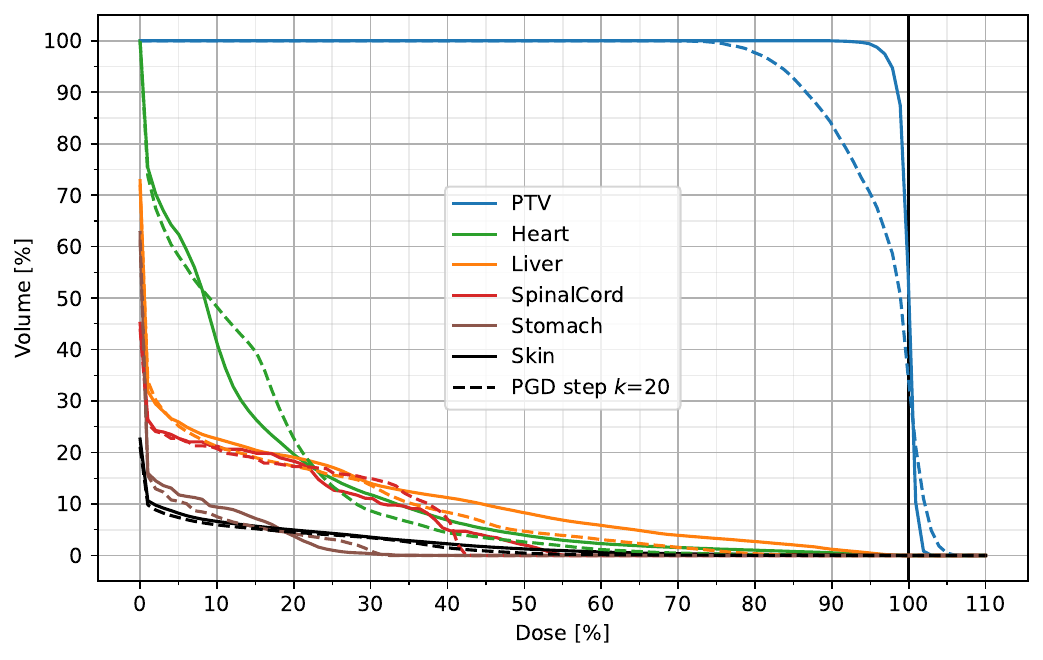}
\hfill
\includegraphics[width=0.49\textwidth]{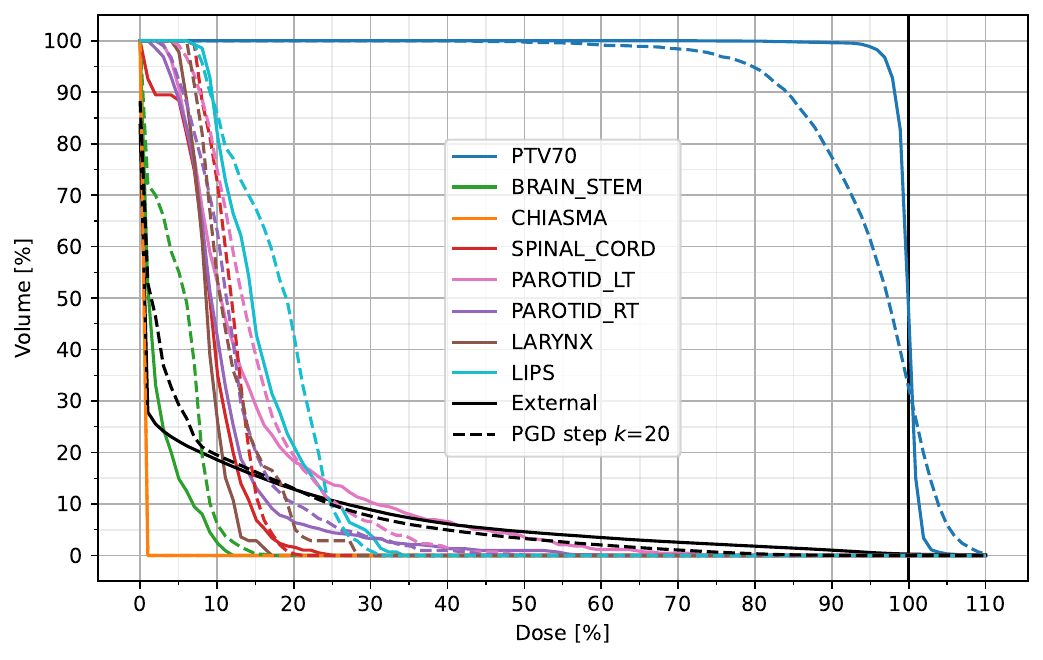}
\caption{Dose-volume histograms for the liver (left) and the head and neck case (right) case of \texttt{full} (solid lines) and the projected gradient descent (PGD) optimization after $k=20$ steps (dashed lines) which we use as a probing function to derive the importance scores.}
\label{app:fig:DVH_PGD}
\end{figure}

First, Figure~\ref{app:fig:samples_rest} depicts the same as Figure~\ref{fig:samples_prostate} but for the liver and head and neck cases.
We can clearly see that body voxels that lie in the beam corridor get a higher probability of being sampled than regular body voxels.
In general, most of the probability mass is placed at OARs and around the PTV. 
\\ 

Second, Figure~\ref{app:fig:DVH_PGD} visualizes the dose-volume histograms as dashed lines for the liver and head and neck cases after $k=20$ steps of projected gradient descent on Equation~\ref{eq:probing}, which we use to derive our importance scores.
Just as in Figure~\ref{fig:DVH} (left), the dose-volume histograms derived from solving the full optimization problem of Equation~\eqref{eq:opt} is depicted as solid lines for reference.
As for the prostate case, the dose-volume histograms obtained from using the solutions $x_{20}$ do not approximate the desired dose-volume histograms sufficiently well.
However, as seen in Figure~\ref{app:fig:samples_rest}, the solutions $x_{20}$ are sufficient to derive meaningful importance scores as seen by the sampling distributions.
\\ 

\begin{figure}[h!]
\includegraphics[width=0.49\textwidth]{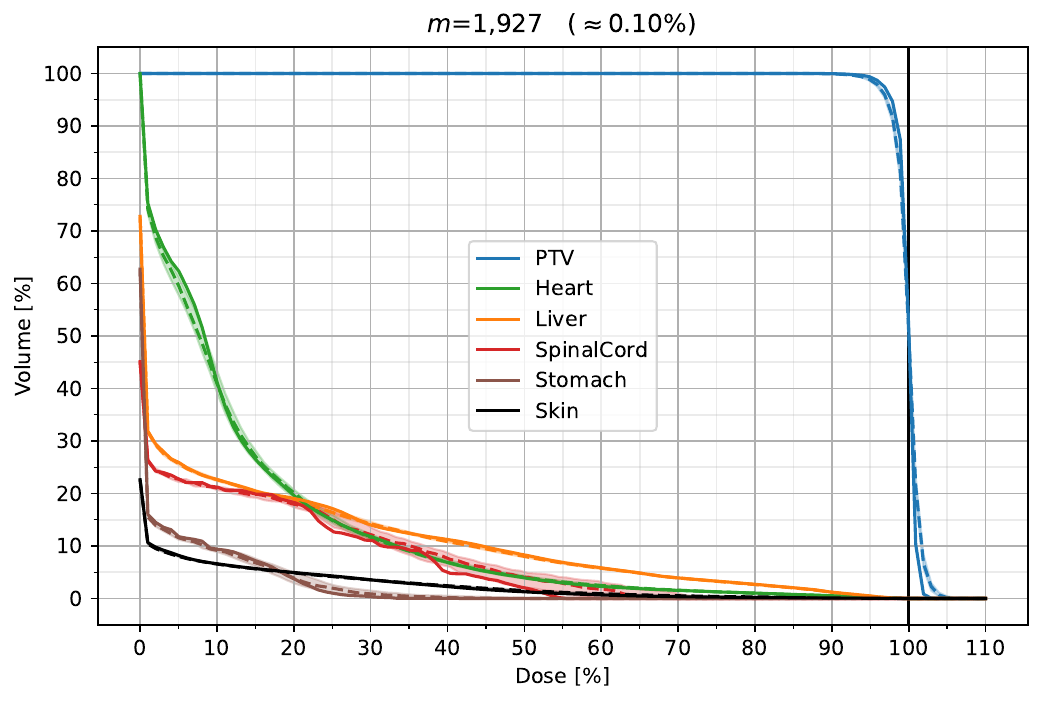}
\includegraphics[width=0.49\textwidth]{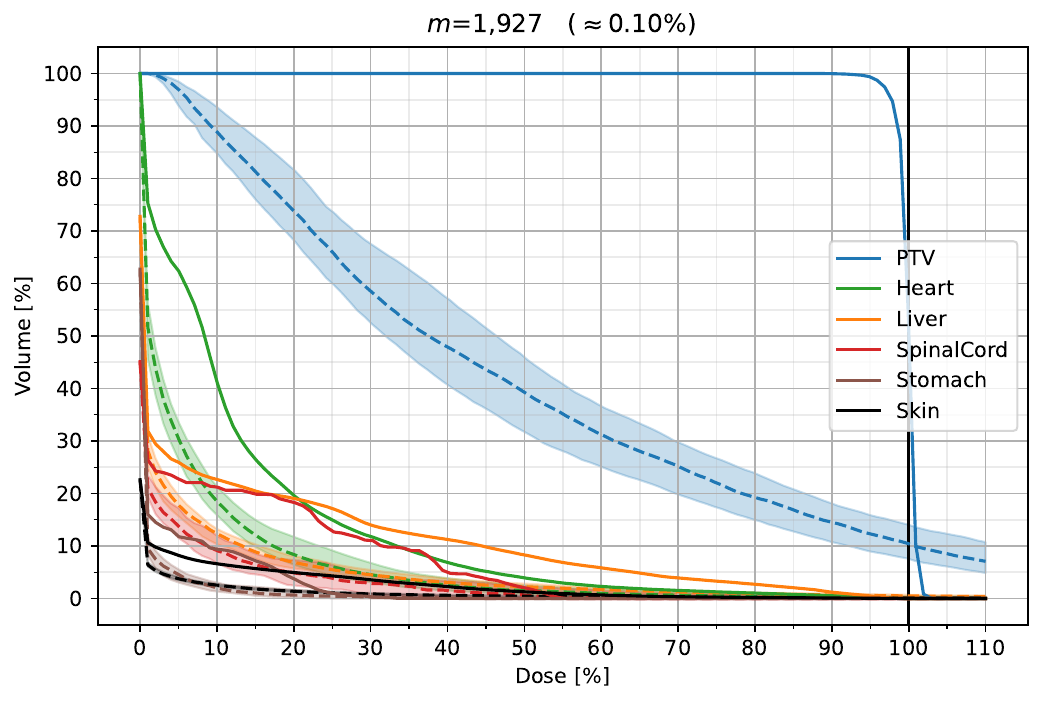}
\includegraphics[width=0.49\textwidth]{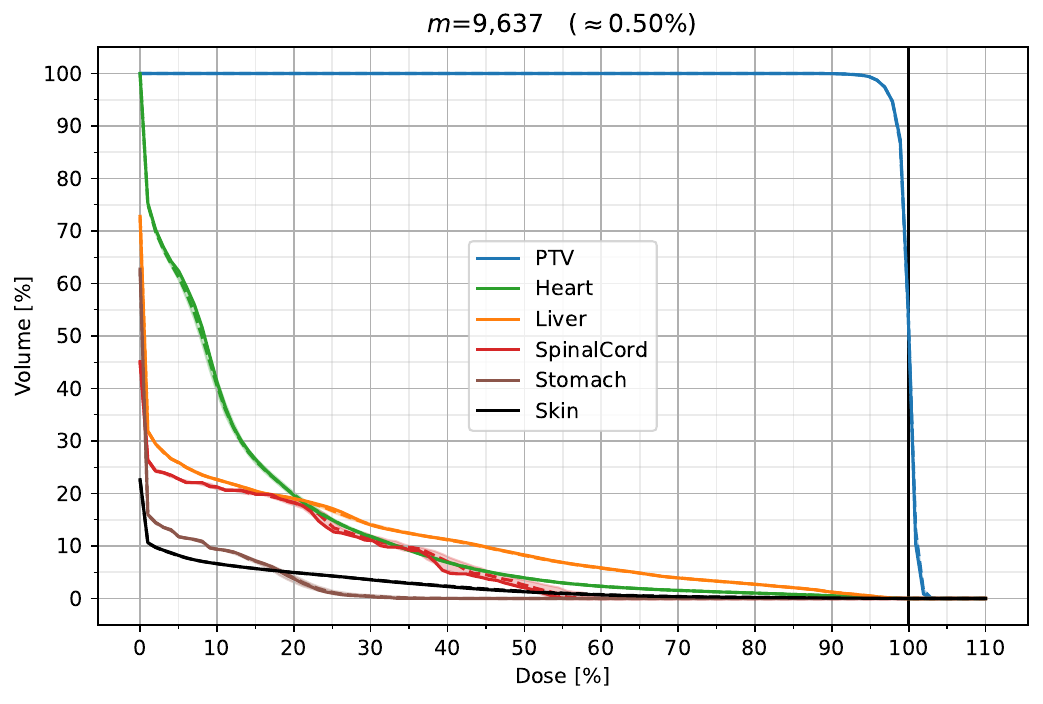}
\includegraphics[width=0.49\textwidth]{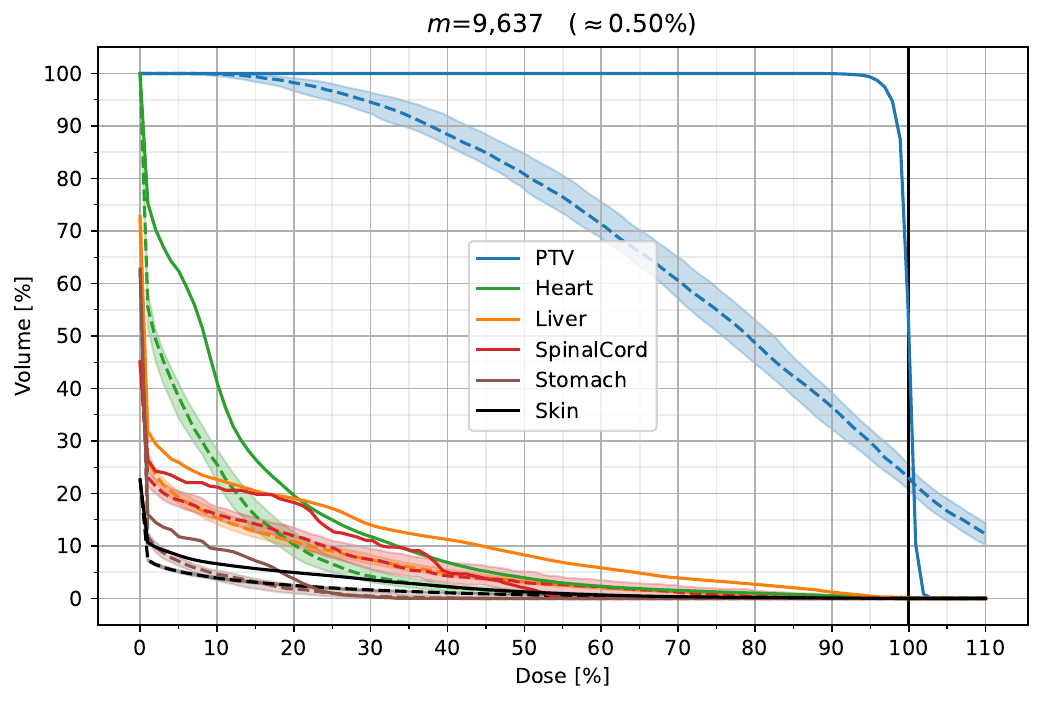}
\includegraphics[width=0.49\textwidth]{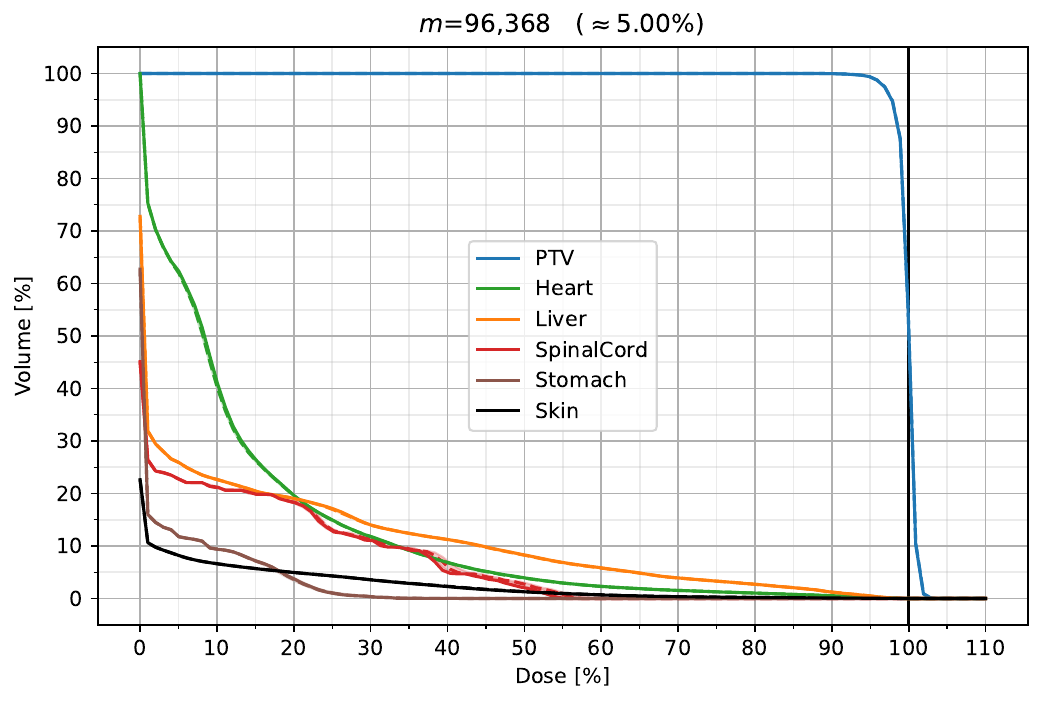}
\includegraphics[width=0.49\textwidth]{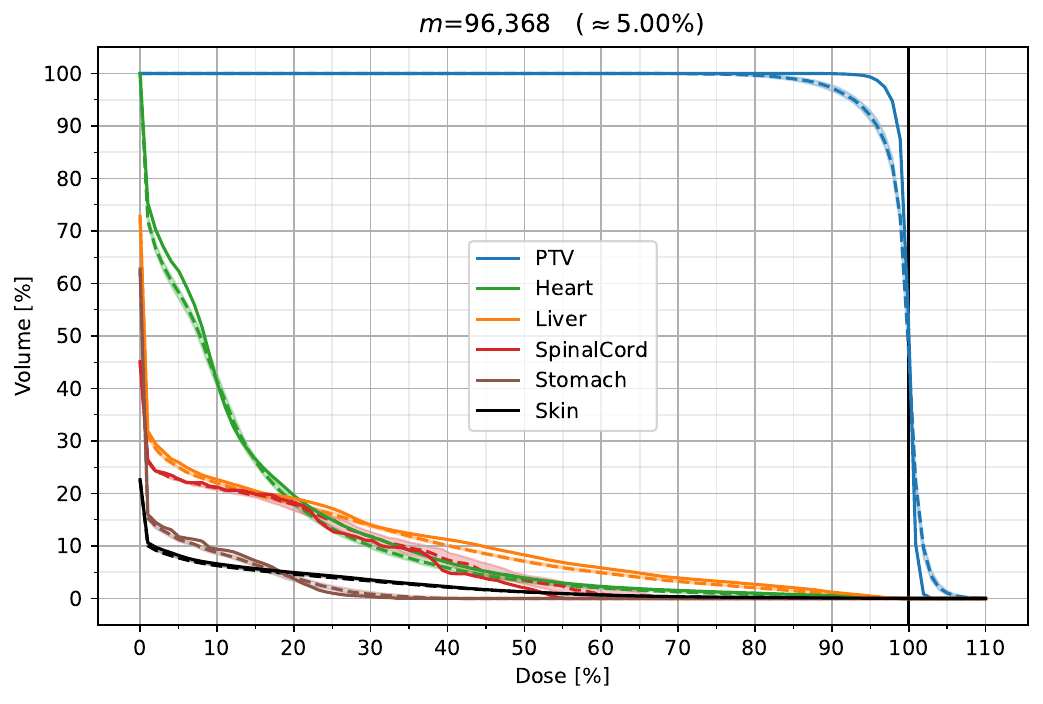}
\includegraphics[width=0.49\textwidth]{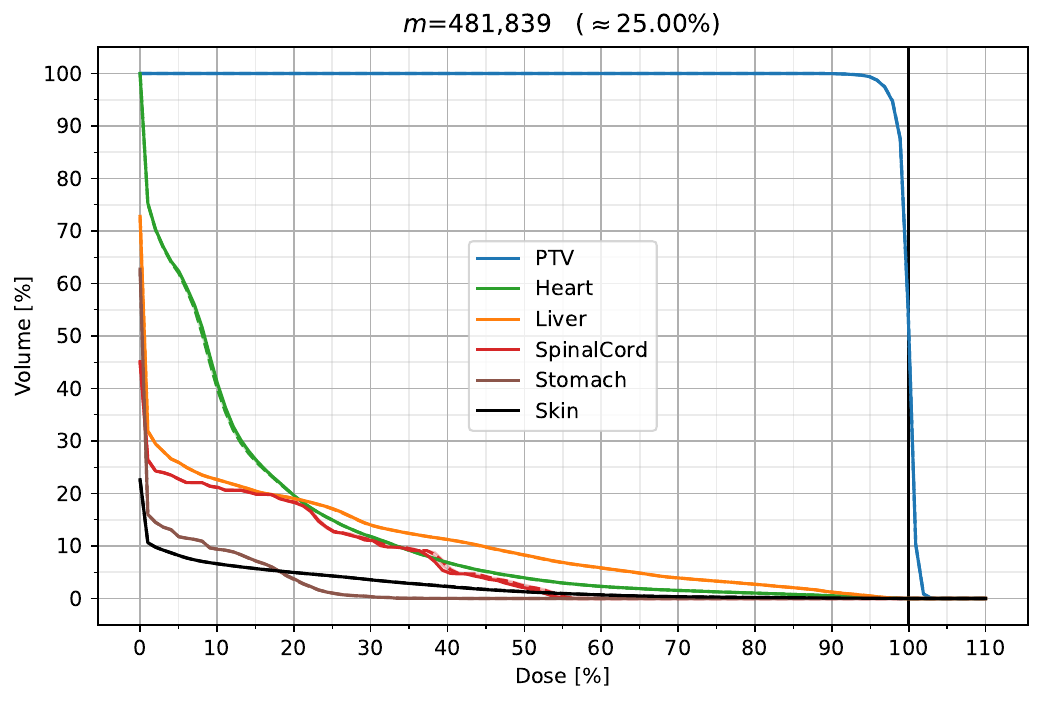}
\includegraphics[width=0.49\textwidth]{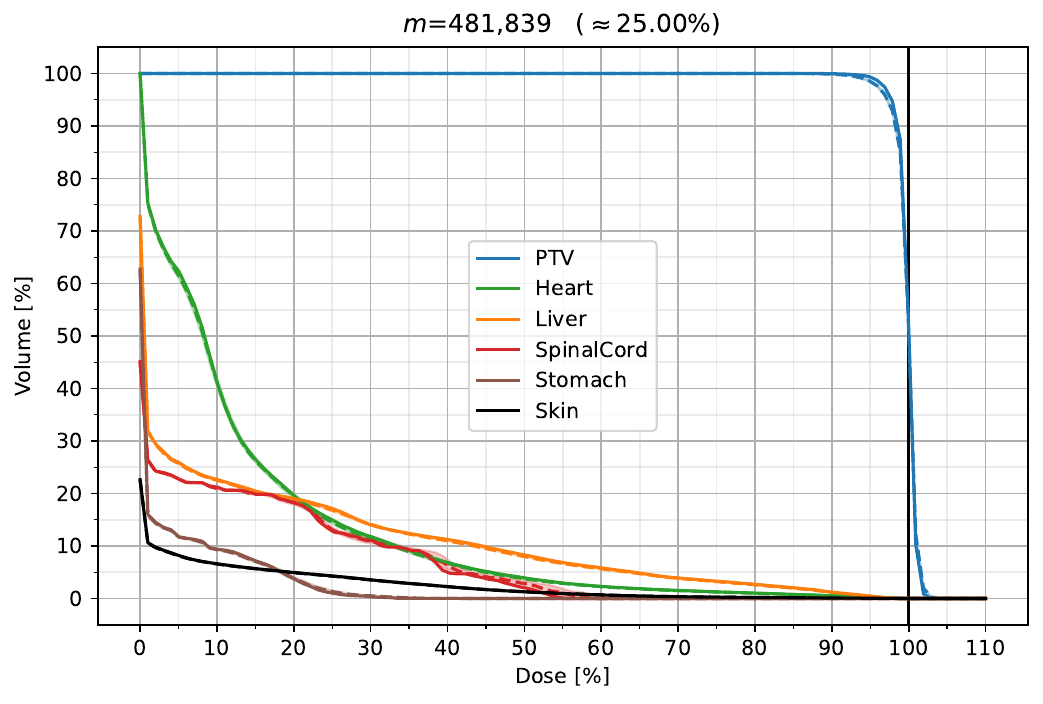}
\caption{Dose-volume histograms for the liver case and $p=2$. The left column depicts the results of our proposed \texttt{gradnorm} subsampling approach while the right column shows \texttt{uniform} subsampling. The rows depict different subsampling sizes~$m$.}
\label{app:fig:DVHdist_liver}
\end{figure}

\begin{figure}
\includegraphics[width=0.49\textwidth]{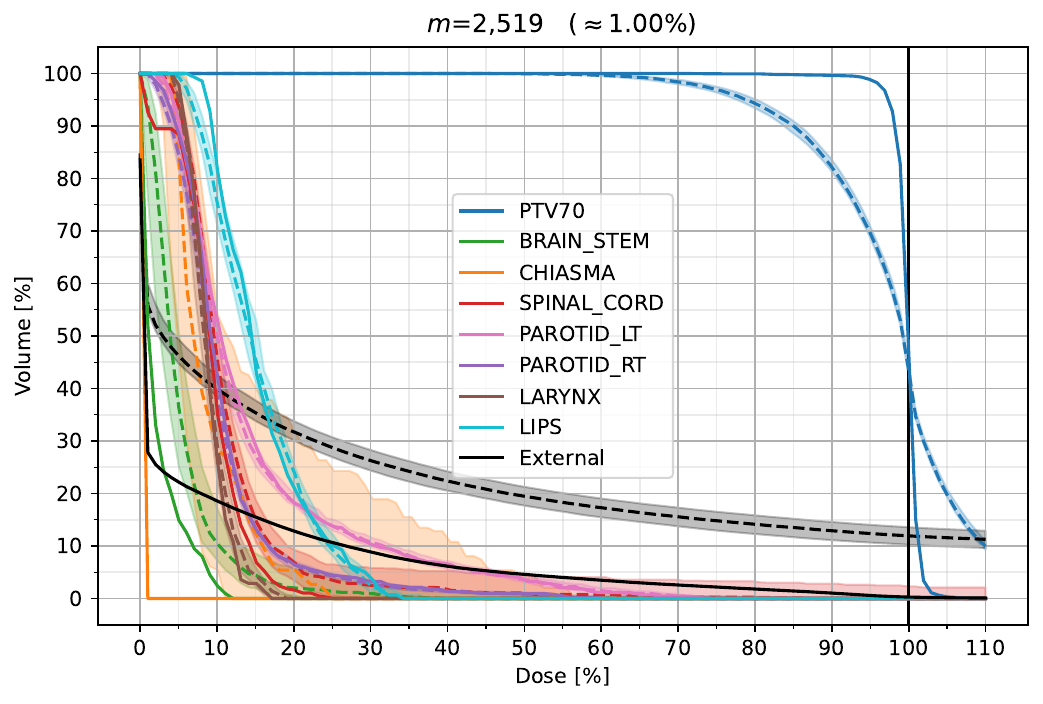}
\includegraphics[width=0.49\textwidth]{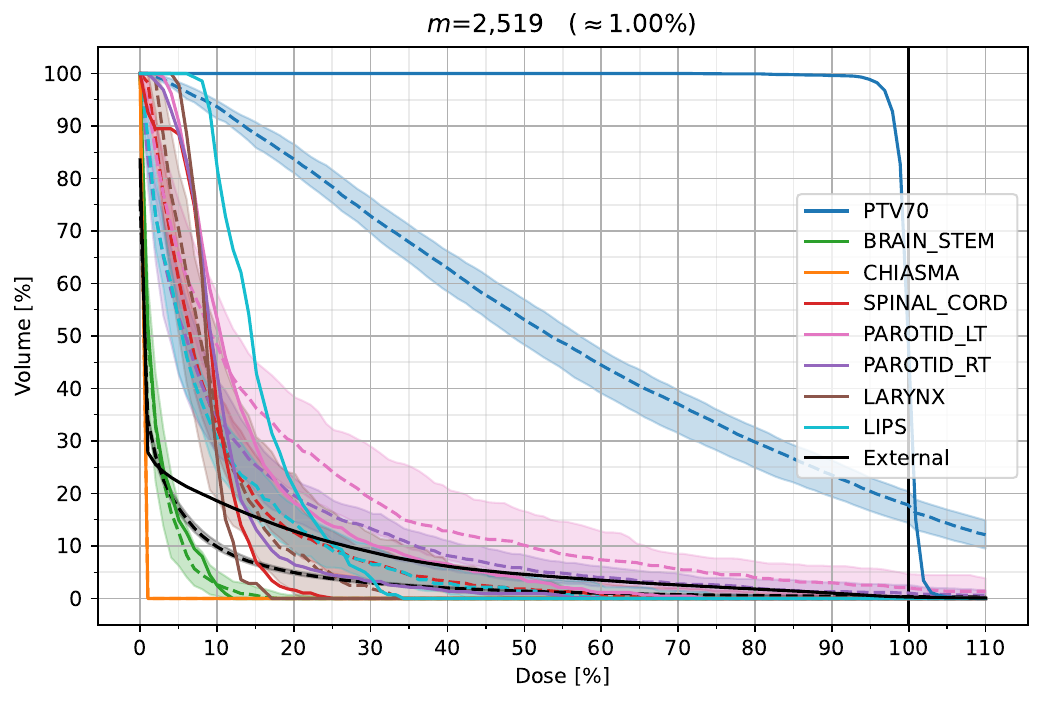}
\includegraphics[width=0.49\textwidth]{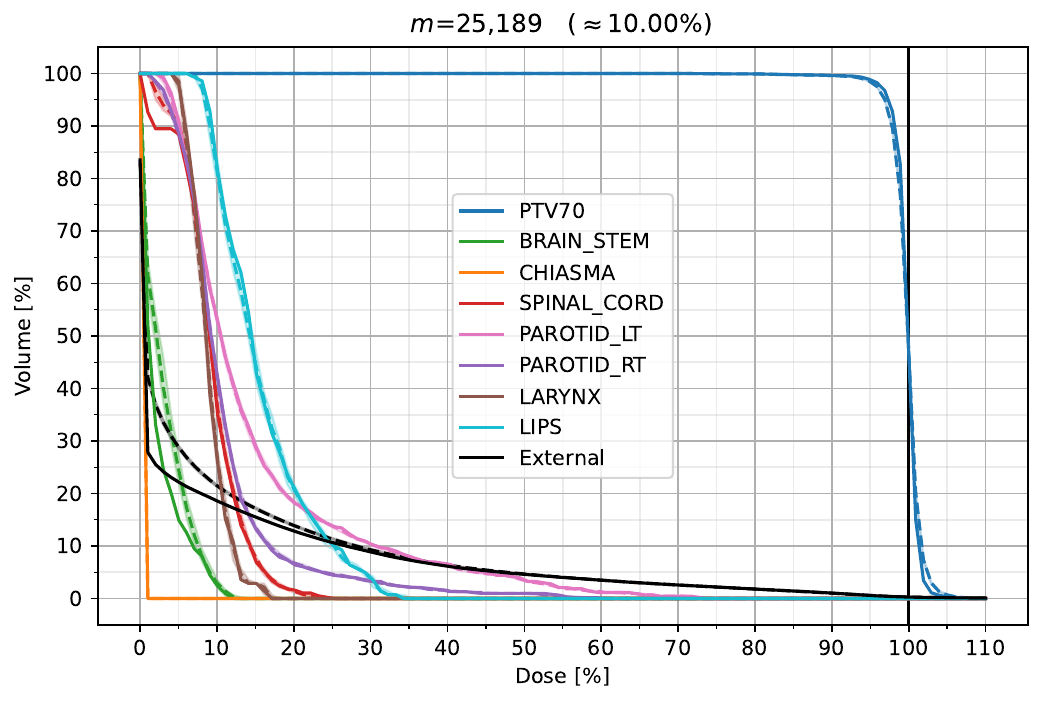}
\includegraphics[width=0.49\textwidth]{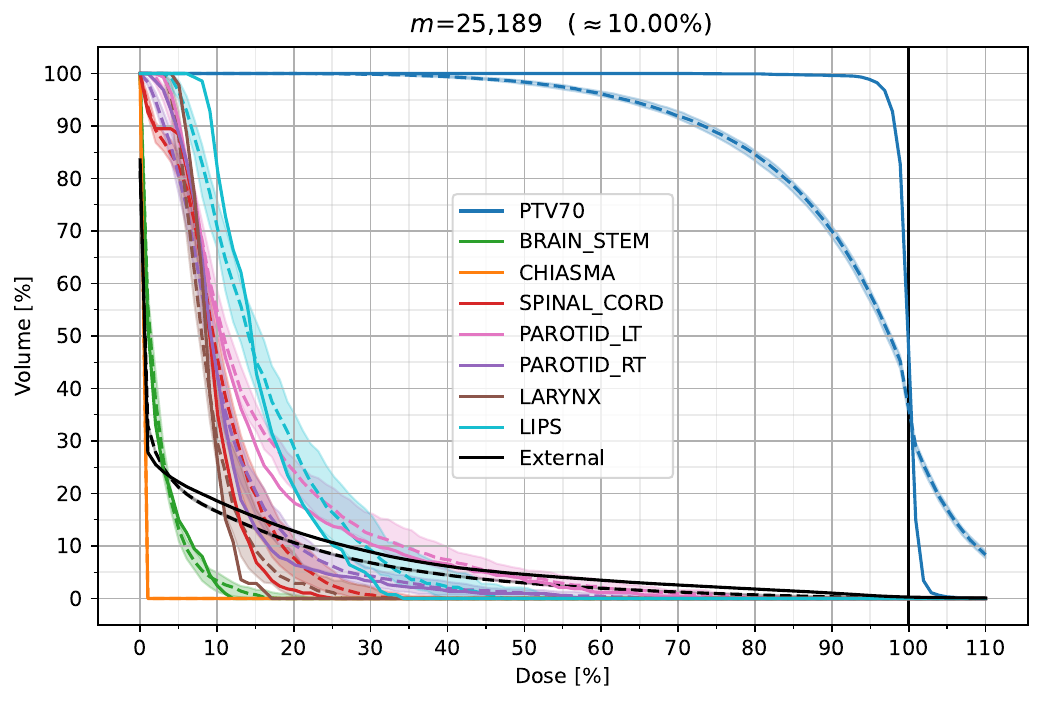}
\includegraphics[width=0.49\textwidth]{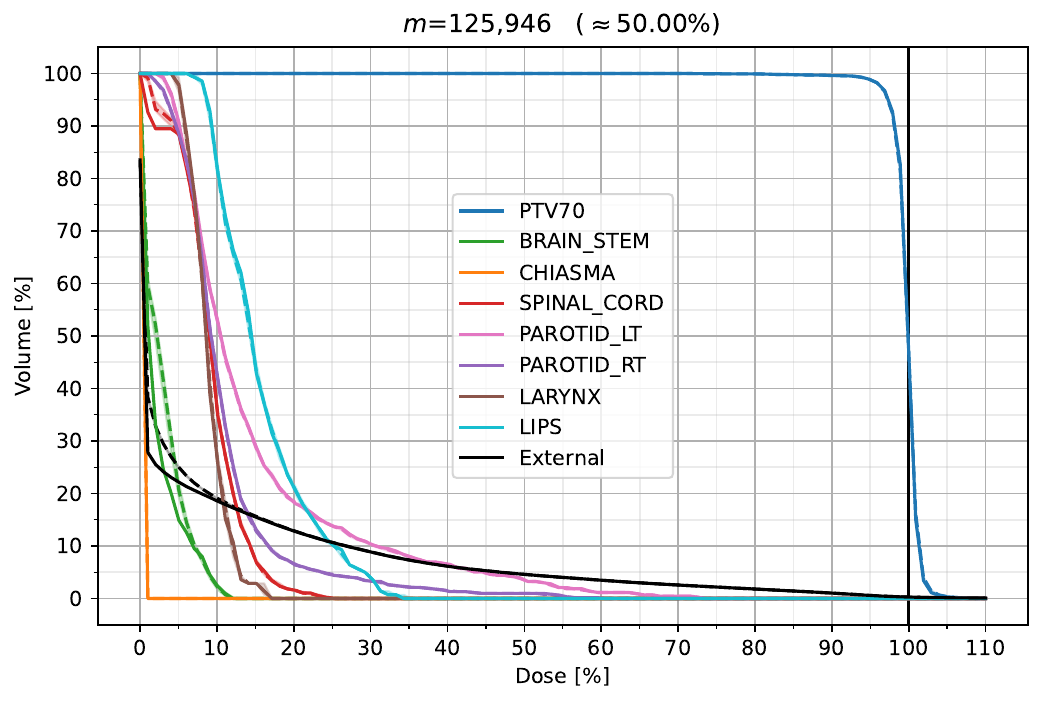}
\includegraphics[width=0.49\textwidth]{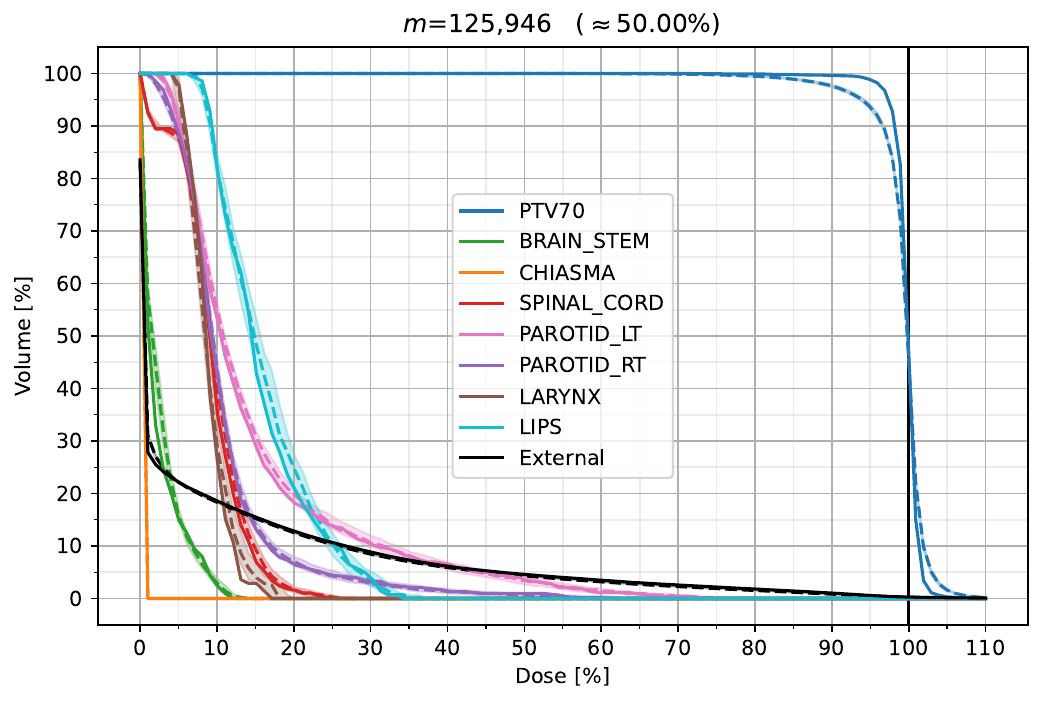}
\includegraphics[width=0.49\textwidth]{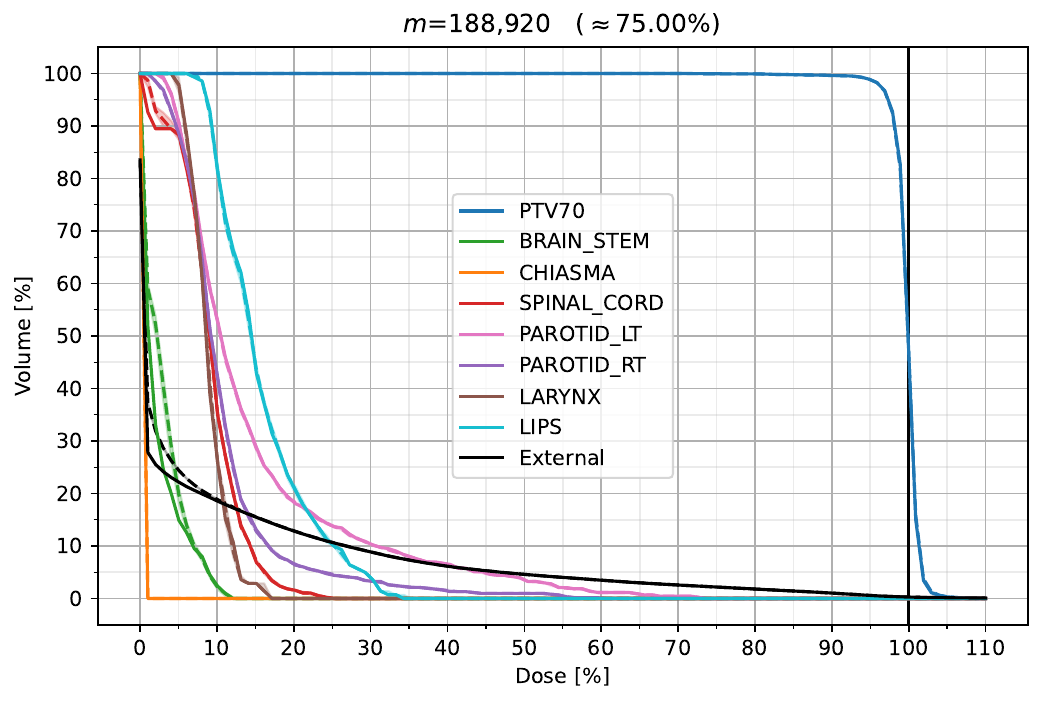}
\includegraphics[width=0.49\textwidth]{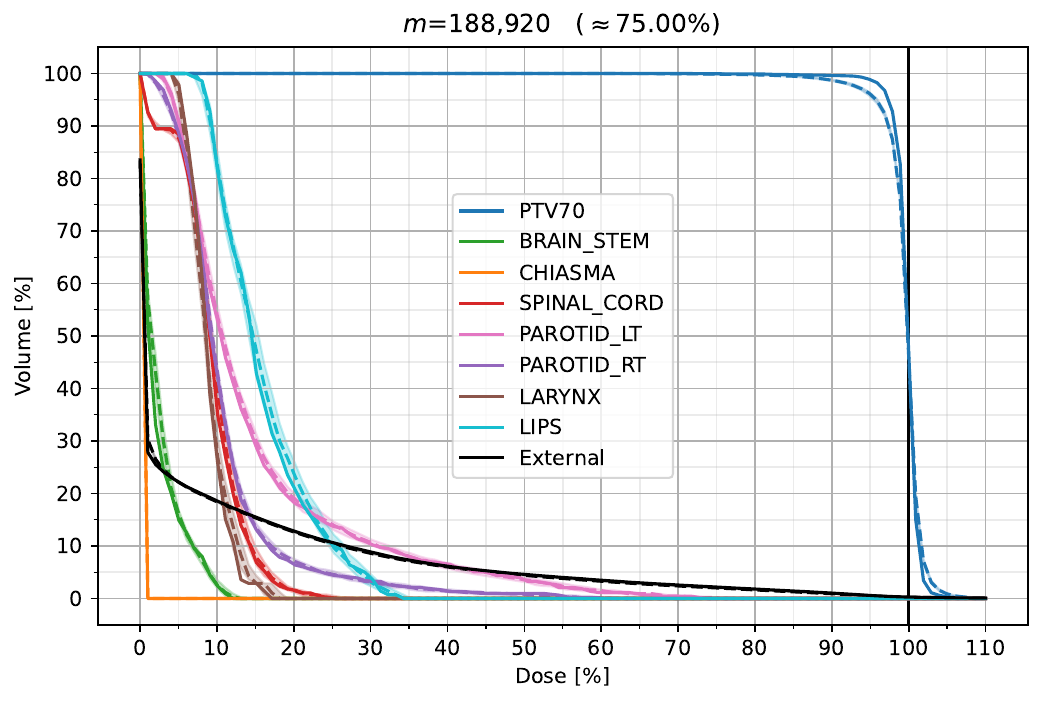}
\caption{Dose-volume histograms for the head and neck case and $p=2$. The left column depicts the results of our proposed \texttt{gradnorm} subsampling approach while the right column shows \texttt{uniform} subsampling. The rows depict different subsampling sizes~$m$.
\looseness=-1}
\label{app:fig:DVHdist_headandneck}
\end{figure}

Next, Figures~\ref{app:fig:DVHdist_liver} and \ref{app:fig:DVHdist_headandneck} show the same scenario as Figure~\ref{fig:DVHdist_prostate} but for the liver and head and neck cases.
The left columns show the performance of our proposed \texttt{gradnorm} score while the right columns depict the performance of a \texttt{uniform} subsample.
Figure~\ref{app:fig:DVHdist_liver} shows that the liver case is especially easy to approximate using \texttt{gradnorm} and a very small subsample of just 0.1\% of the total voxel amount.
Contrary, \texttt{uniform} needs at least 25\% to achieve a similar performance.
The scenario is slightly different for the head and neck case as visualized in Figure~\ref{app:fig:DVHdist_headandneck} where as much as 10\% of the voxels are needed to achieve an acceptable approximation.
\\

\begin{figure}[t]
\centering
\includegraphics[width=0.49\textwidth]{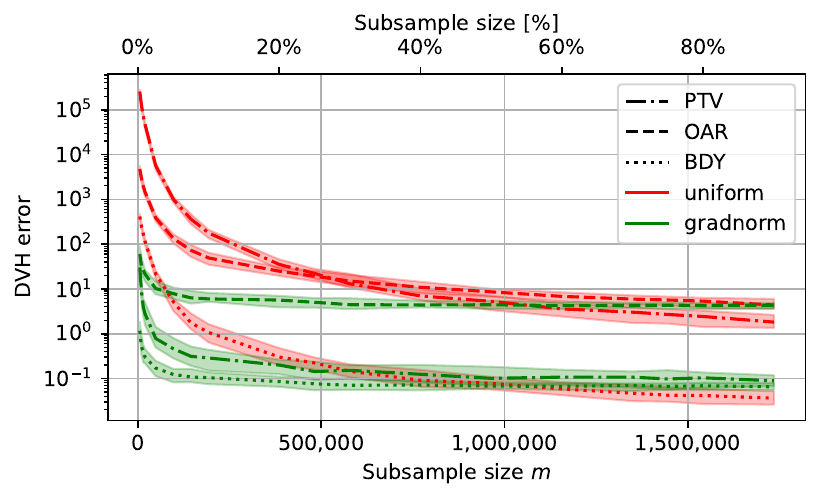}
\hfill
\includegraphics[width=0.49\textwidth]{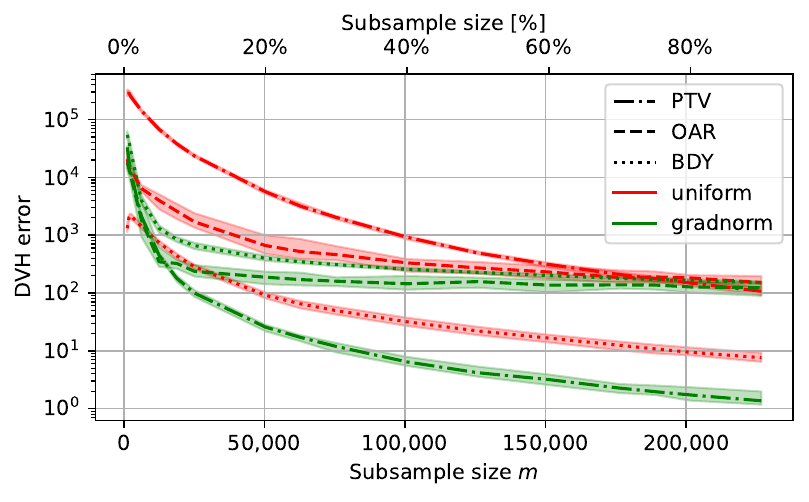}
\caption{Error of dose-volume histogram approximations of \texttt{uniform} and \texttt{gradnorm} when compared to \texttt{full} for the liver (left) and the head and neck (right) case. The error is split among the structures $\Ss=\{\text{PTV}, \text{OAR}, \text{BDY}\}$.}
\label{app:fig:DVHerror_susample}
\end{figure}

\begin{figure}[t]
\centering
\includegraphics[width=0.49\textwidth]{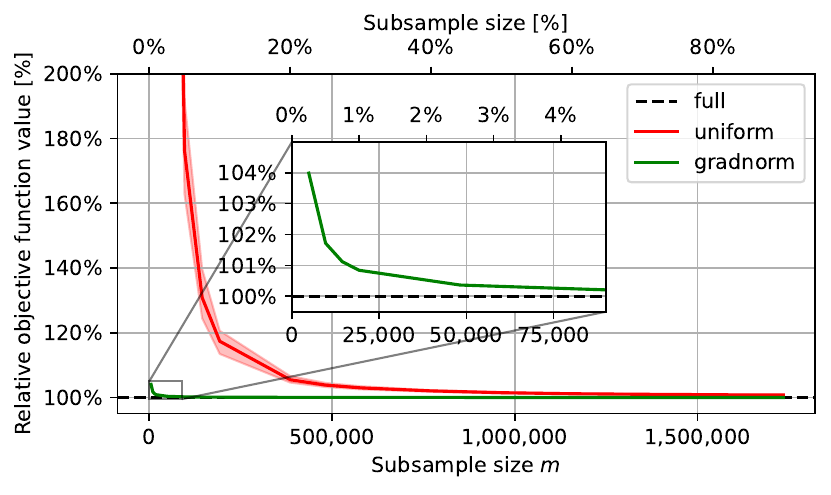}
\hfill
\includegraphics[width=0.49\textwidth]{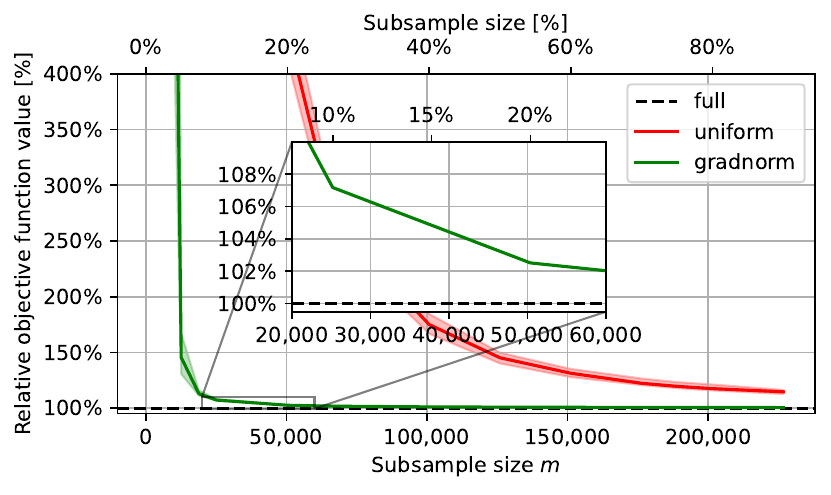}
\caption{Relative objective function values of \texttt{uniform} and \texttt{gradnorm} when compared to \texttt{full} for the liver (left) and the head and neck (right) case.}
\label{app:fig:relobj_subsample}
\end{figure}

Figure~\ref{app:fig:DVHerror_susample} depicts the error of dose-volume histogram approximations and we can see that the only structure on which \texttt{uniform} yields a better approximation than \texttt{gradnorm} is BDY.
For the liver case, this is only true for large sample sizes but for the head and neck case, this effect is visible much earlier.
This can be also seen in Figure~\ref{app:fig:DVHdist_headandneck} where the black line representing the BDY structure (named External) struggles to approximate the low dose region.
However, we note that -- in turn -- the more important PTV and OAR structures are approximated sufficiently well.

The relative objective function approximations of \texttt{uniform} and \texttt{gradnorm} are shown in Figure~\ref{app:fig:relobj_subsample} for the liver (left) and head and neck (right) case, respectively.
For the liver case, \texttt{gradnorm} yields an error of less than 1\% when using more than 1\% of all voxels whereas \texttt{uniform} needs significantly more voxels to achieve a similar approximation.
As mentioned above, the head and neck case appears to be more difficult and Figure~\ref{app:fig:DVHdist_headandneck} suggests that \texttt{gradnorm} needs at least 10\% of the voxels.
Now, Figure~\ref{app:fig:relobj_subsample} (right) reveals that this choice yields an approximation of the objective function that only deviates 7\% of the objective function value obtained from the \texttt{full} solution.
\\ 

\begin{figure}[t]
\centering
\includegraphics[width=0.49\textwidth]{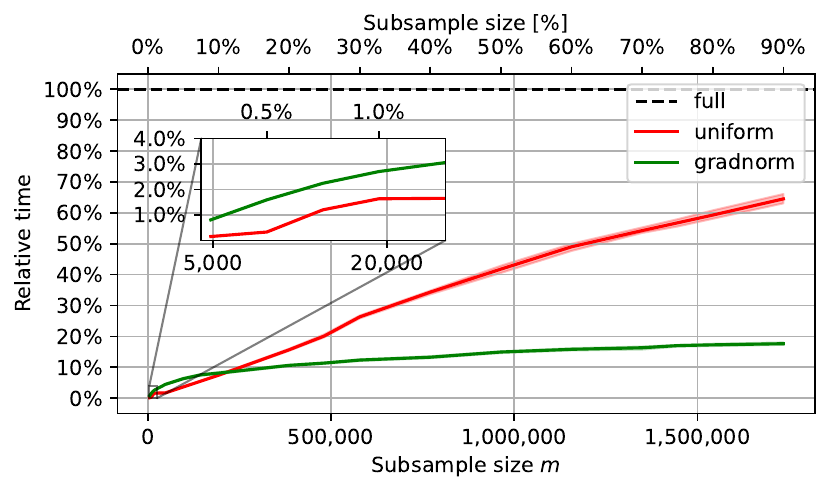}
\hfill
\includegraphics[width=0.49\textwidth]{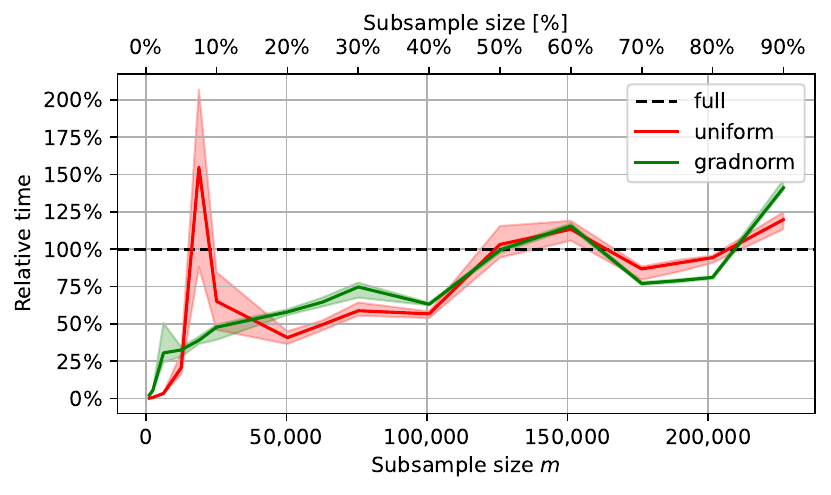}
\caption{Computation time of \texttt{uniform} and \texttt{gradnorm} relative to \texttt{full} for the liver (left) and the head and neck (right) case.}
\label{app:fig:reltime_subsample}
\end{figure}

Finally, Figure~\ref{app:fig:reltime_subsample} depicts the relative computation times of solving the reduced optimization problems when compared to the time it takes to solve the corresponding full problem.
For the liver case on the left-hand side, subsampling only 1\% of all voxels using \texttt{gradnorm} lowers the computation times to less than 3\% of the time \texttt{full} needs on all voxels.
Whereas the optimization times are rather stable for the prostate and liver cases, the situation is -- once again -- different for the head and neck case.
However, using the proposed 10\% of all voxels using \texttt{gradnorm} allows us to reduce the computation time to less than 50\%.

\clearpage 

\bibliography{literature}
\bibliographystyle{dcu}

\end{document}